\documentclass[twocolumn]{emulateapj}

\usepackage{psfig}
\usepackage{color}
\usepackage{natbib}
\usepackage{ulem}
\usepackage{soul}

\def\thesource{{3C~454.3}}

\def\mgii{{Mg~{\sc II}}}
\def\smarts{\textit{SMARTS}}
\def\fermi{\textit{Fermi}}

\def\lblr{L$_{BLR}$}

\def\g{$\gamma$}
\def\bet{$\beta$}
\newcommand{\ang}{\mbox{$\:$\AA}}

\slugcomment{Accepted to the Astrophysical Journal}

\shortauthors{Isler et al.}
\shorttitle{Time Resolved BLR Behavior of \thesource}

\begin{document}

\title{A Time Resolved Study of the Broad Line Region in Blazar 3C 454.3}

\author{Jedidah C. Isler\altaffilmark{1,2}, C.M. Urry\altaffilmark{1,3}, P. Coppi\altaffilmark{1},\\
C. Bailyn\altaffilmark{1}, 
R. Chatterjee\altaffilmark{4},
G. Fossati\altaffilmark{5}, 
E.~W. Bonning\altaffilmark{6},
L. Maraschi\altaffilmark{7}, 
M. Buxton\altaffilmark{1} 
}

\altaffiltext{1}{Department of Astronomy, Yale University, PO Box 208101, New Haven, CT 06520-8101; jedidah.isler@yale.edu}
\altaffiltext{2}{National Research Council Ford Dissertation Fellow, National Science Foundation, Graduate Research Fellow, NASA-Jenkins Graduate Fellow}
\altaffiltext{3}{Department of Physics and Yale Center for Astronomy
and Astrophysics, Yale University, PO Box 208121, New Haven, CT 06520-8120}
\altaffiltext{4}{Department of Physics, Presidency University, 86/1 College Street, Kolkata-700073, WB, India}
\altaffiltext{5}{Department of Physics and Astronomy, Rice University, Houston, TX 77005}
\altaffiltext{6}{Quest University Canada, 3200 University Boulevard, Squamish, BC, V8B 0N8, Canada}
\altaffiltext{7}{INAF - Osservatorio Astronomico di Brera, V. Brera 28, I-20100 Milano, Italy}

\begin{abstract}
We present multi-epoch optical observations of the blazar \thesource\ (z = 0.859) from 2008 August through 2011 December, using the \smarts\ Consortium 1.5m+RCSpectrograph and 1.3m+ANDICAM in Cerro Tololo, Chile. 
The spectra reveal that the broad optical emission lines \mgii, H\bet\
and H\g\ are far less variable than the optical or \g-ray continuum.  
Although, the \g-rays varied by a factor of 100 above the EGRET era flux, the lines generally vary by a factor of 2 or less. Smaller variations in the \g-ray flux did not produce significant variation in any of the observed emission lines.
Therefore, to first order, the ionizing flux from the disk changes only slowly during large variations of the jet. 
However, two exceptions in the response of the broad emission lines are reported during the largest \g-ray flares in 2009 December and 2010 November, when significant deviations from the mean line flux in H\g\ and \mgii\ were observed. H\g\ showed a maximum 3$\sigma$ and 4$\sigma$ deviation in each flare, respectively, corresponding to a factor of 1.7 and 2.5 increase in flux.
\mgii\ showed a 2$\sigma$ deviation in both flares; no variation was detected in H\bet\ during either flare. 
These significant deviations from the mean line flux also coincide with 7mm core ejections reported previously \citep{Jorstad12}.
The correlation of the increased emission line flux with mm core ejections, and \g-ray, optical and UV flares suggests that the broad line region extends beyond the \g-emitting region during the 2009 and 2010 flares.
\end{abstract}


\section{INTRODUCTION}
\label{sec:intro}
Active galactic nuclei (AGN) are actively accreting supermassive black holes at the centers of massive galaxies. 
They have been well studied since their discovery 50 years ago \citep[e.g.][etc.]{Antonucci84, Peterson86, Tadhunter92, Maraschi92, Ghisellini93, Urry95,Korista97, Fossati98, Kaspi05, Marscher08}, and in broad terms, their basic characteristics are reasonably well known. 
An ultraviolet-X-ray continuum is produced near the black hole, probably in an accretion disk and energetic corona.
Optical and near-infrared spectra of AGN reveal emission lines produced in gas clouds photoionized by this continuum.
In approximately 10\% of the total AGN population, a jet of radio-emitting plasma moves outward relativistically \citep{Kellermann89, Urry95}. We still do not know how jets form and are collimated, what distribution of jet energies nature produces, or how (if) the accretion flow is linked to the jet.

Blazars are AGN oriented with the jet axis at very small angles to the line of sight.
This orientation causes Doppler beaming of the observed jet luminosity, making the jet appear brighter and the variability timescale shorter.
Beaming makes blazars an ideal laboratory for studying the relativistic jets, because the observed jet emission dominates the  thermal emission.
The observed jet luminosity, L, is related to the intrinsic jet luminosity, $\mathcal L$, by the Doppler beaming factor, $\delta$: L=$\delta^{4}\mathcal L$ (with luminosity integrated over energy). 
For blazars with orientation angles that are small with respect to the line of sight, $\delta\sim\Gamma$ at $\theta$=$\Gamma^{-1}$ and $\beta_{app}\sim\Gamma\beta$, where $\Gamma$ is the bulk Lorentz factor, $\beta_{app}$ is the observed velocity of the emitter, and $\beta$ is the true bulk velocity. 
Typically, $\Gamma$ $\sim$ $\delta$ $\sim$10 in blazars.

Outflow energy is thought to be extracted from the black hole through the accretion disk or black hole spin \citep{BlandfordPayne82, BlandfordZnajek77}. 
Thus, the relationship between the accretion disk and relativistic jet is of primary importance to understanding the energy budget of the system. 
The accretion disk light is often swamped by the non-thermal jet emission however, even in the optical regime. 
Still, in some blazars we can see (unbeamed) emission lines and even a hint of a thermal accretion disk spectrum \citep{Pian98, Raiteri07_0235, Bonning09, DAmmando09, Bonnoli11, Raiteri11}, polarization data provides an independent detection of the Big Blue Bump in blazars \citep{Smith86, Smith88}.

Perhaps the most powerful tool to address the connection between the disk and jet is time variability. By monitoring the different components and emission regimes, and tracking the leads and lags between them, one can in principle build up a full geometric picture of the flow of matter and resulting radiation. 

The short timescale \g-ray variability of blazars can now be studied extremely well thanks to the launch of the \fermi\  \g-ray satellite in 2008 June. 
Since then, several groups have been monitoring \g-ray bright blazars in other wavebands.
In particular, \smarts\ daily optical-infrared photometry and bi-monthly optical spectroscopy were obtained for several southern hemisphere sources \citep{Bonning12}, including \thesource\ \citep{Bonning09}.  This blazar was the brightest \g-ray source in the \fermi\ sky during and after launch. During the EGRET era it was much fainter, with a maximum jet flaring flux(E$>$ 100MeV) of F$_{100}\sim$ 0.5$\times$10$^{-6}$ photons cm$^{-2}$ s$^{-1}$, more than a factor of 120 times fainter than the brightest state during the \fermi\ era.

\begin{figure*}
\plotone{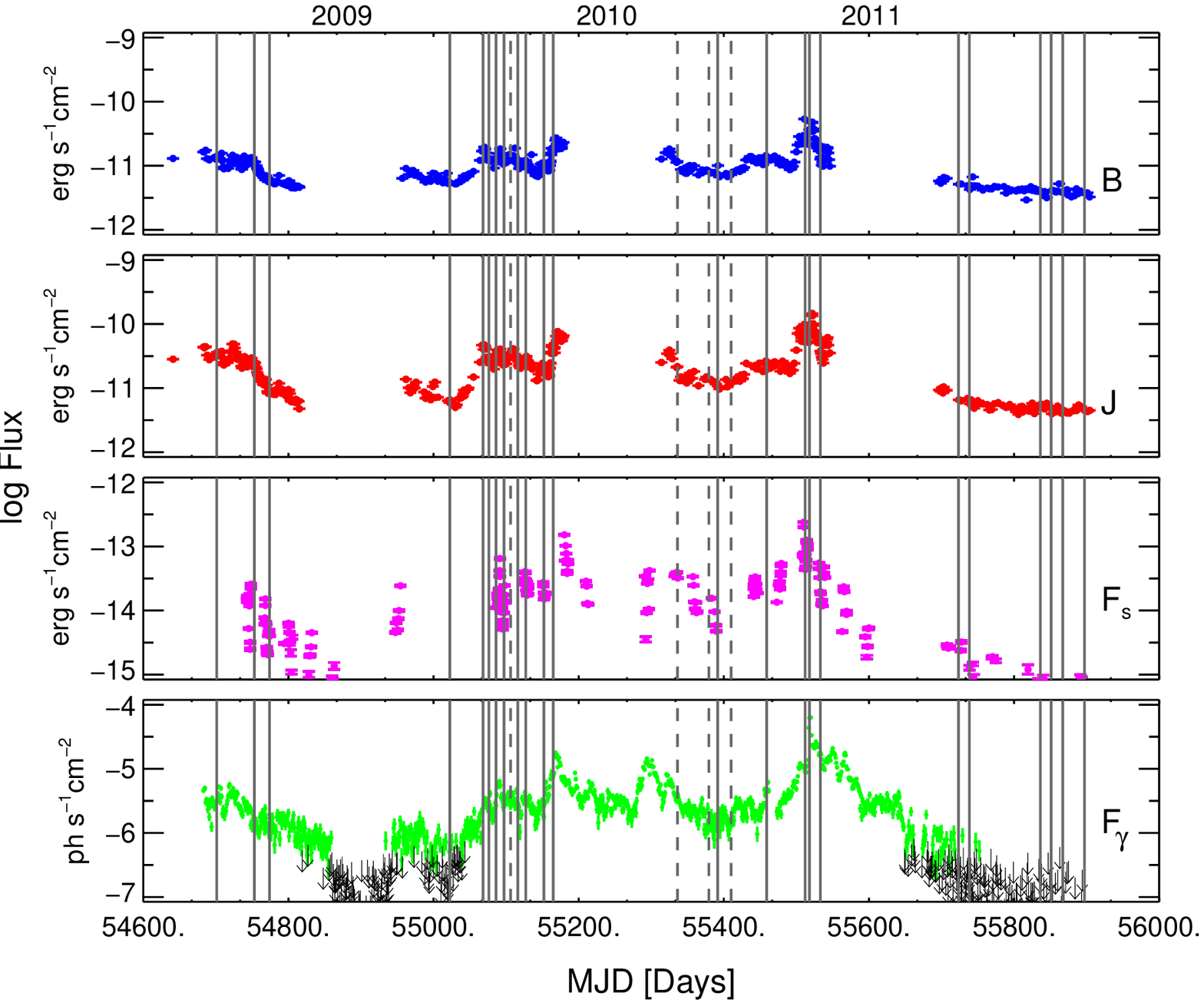}
\figcaption{The 3.3 year light curves of \thesource\ from \smarts\ B- \& J- band, polarized optical flux from Steward monitoring (F$_s$), and Fermi (E$>$ 100MeV) \g-rays, in log flux units.  Vertical lines mark dates of optical spectroscopy, with dashed lines indicate dates for which \smarts\ spectroscopy and photometry were taken on separate nights. The optical spectroscopy samples periods of high, moderate and low \g-ray activity over the duration of monitoring. The \g-ray fluxes with low significance (TS$<$25) are plotted as upper limits. We note the low polarized flux observed near MJD 55700 that also corresponds to low \fermi\ \g-ray fluxes.  \label{fig:lc}}
\end{figure*}

Optical-infrared photometry of \thesource\ during the first year of \fermi\ operations showed correlated variability with \g-ray and ultraviolet light curves \citep[Fig \ref{fig:lc}]{Bonning09}, a result that was confirmed for subsequent flares in the same source \citep{Pacciani10, Raiteri11, Vercellone11, Jorstad10}. 
Five major \g-ray flares have been recorded for \thesource\ over the three year period discussed here. 
High resolution observations of the radio core of \thesource\ have also been undertaken \citep{Jorstad05, Jorstad10,Jorstad12}. 
Recent observations show the 1mm and 7mm radio core emission is well correlated with the \g-ray and optical flaring events, with 7mm core ejections detected in the 2009 December and 2010 November flaring periods \citep{Jorstad12}. 
This correlation suggests that in addition to the synchrotron-emitting electrons and upscattered \g-ray radiation being cospatial \citep{Bonning09}, the high-frequency radio core emission must also be in close proximity given their simultaneous flaring episodes. 

Emission line variability information in blazars comparable to the reverberation mapping studies of moderate luminosity AGN is scarce \citep{Peterson86, Clavel91, Netzer97}, although some line variability  has been observed on timescales of months \citep{Zheng86, Bregman86, Ulrich97, Perez89, Corbett00}. 
For two prominent blazars, the continuum variations are a factor of 2 (3C273) to a factor of 50 (3C279) times larger in the optical/ultraviolet than the emission line variations, which are less than 25\% in amplitude
\citep{Ulrich93,Falomo94,  Koratkar98}. 
This lack of large amplitude line variability in blazars implies the underlying photoionizing flux comes from a relatively constant source, presumably the accretion disk. 
Historically, no evidence for an increase in line flux related to the jet flares was detected, even when the beamed contribution to the UV continuum swamped the unbeamed thermal disk contribution at periods of high jet activity \citep[e.g.,][]{Smith11}. 

However, the lack of long-term simultaneous high-energy data and multi-epoch emission line studies made it difficult to detect any disk-jet connection.
Multi-epoch optical spectroscopy of mostly northern hemisphere sources was carried out in conjunction with \fermi\ observations, including \thesource\ \citep{Smith09, Benitez10}. Recently, a 40\% variation in the \mgii\ line was reported \citep{LT13}.
Other contemporaneous, spectroscopic studies have compared single-epoch \fermi-detected blazar emission lines and the total radiative \g-ray luminosity to derive correlations between the broad line region and \g-ray luminosity \citep{Chen09, Ghisellini11, Sbarrato12, Shaw12}. 

In this paper we present time-resolved spectroscopy of the bright, superluminal blazar \thesource\ obtained with \smarts, which we supplement with publicly available data. The variability of the strongest features, \mgii, H\bet, and H\g\ emission lines, is discussed.  We then compare to the continuum variability, in particular the optical and \g-ray flux, which comes from the aligned jet. 
In \S \ref{sec:obs} we describe the \smarts\ observational program and \fermi\ data. In \S \ref{sec:results}, we discuss the line variations.
We discuss implications of our findings in \S \ref{sec:disc} and conclusions in \S \ref{sec:sum}. We use concordance cosmology to describe our results: H$_o$ = 70 km s$^{-1}$ Mpc$^{-1}$, $\Omega_m$=0.27, $\Omega_\Lambda$=0.73.

\begin{deluxetable*}{cclrcrccrrrrrrr}
\tablecaption{\smarts\ 1.5m Observation Log}

\tablehead{\colhead{UTC} & \colhead{MJD} & \colhead{V} & \colhead{$\sigma_V$} &  \colhead{R} & \colhead{$\sigma_R$}  & \colhead{EW (\mgii)} & \colhead{$\sigma_M$} & \colhead{EW (H$\beta$) } & \colhead{$\sigma_\beta$} & \colhead{EW( H\g) } & \colhead{$\sigma_\gamma$}\\
\colhead{[YYYYMMDD]}&\colhead{} &\colhead{[mag]} &\colhead{[mag]}& \colhead{[mag]} & \colhead{[mag]} &\colhead{[\ang]} & \colhead{[\ang]}&\colhead{[\ang]}& \colhead{[\ang]}&\colhead{[\ang]}& \colhead{[\ang]}} 

\startdata
20080823 & 54701.2 & 15.094 & 0.007 &  14.579 & 0.006& 2.75 & 1.12 & 2.66 & 1.72 & 3.39 & 2.26   \\
 20081014 & 54753.1 & 15.353 & 0.013 &  14.871 & 0.011  & 4.57 & 1.97 & 3.76 & 2.69 & 6.19 & 4.98  \\
 20081104 & 54774.0 & 16.145 & 0.011 &  15.649 & 0.01 & 8.22 & 1.69 & 12.11 & 5.24 & 13.48 & 7.34  \\
  20090710 & 55022.4 & 16.231 & 0.02 &  15.868 & 0.018 & 8.13& 1.70 & 13.02 & 6.02 & 13.95 & 7.28  \\
  20090825 & 55068.2 & 14.904 & 0.006 & 14.389 & 0.005 & 2.75 & 0.36 & 4.23 & 1.04 & 3.04 & 0.51 \\
  20090902 & 55076.2 & 15.214 & 0.01 &  14.736 & 0.008 & 4.29 & 0.98 & 5.70 & 1.83 & 3.60 & 0.79  \\
  20090912 & 55086.1 & 15.309 & 0.007 &  14.816 & 0.006 & 3.94 & 0.51 & 4.57 & 1.06 & 2.93 & 0.55  \\
  20090923 & 55097.1 & 15.435 & 0.007&  14.911 & 0.006  & 2.11 & 0.35 & 4.61 & 1.36& 3.68 & 1.21  \\
  20091002$^\dag$ & 55106.2 & 15.147 & 0.011 &  14.707 & 0.009 & 3.77 & 0.86 & 4.71 & 1.49 & 2.25 & 0.83  \\
  20091012 & 55116.1 & 15.417 & 0.011 & 14.911 & 0.006 & 4.03 & 0.51 & 5.54 & 1.24 & 4.26 & 0.64  \\
  20091023 & 55127.1 & 15.319 & 0.008 & 14.816 & 0.006 & 4.01 & 0.57 & 5.78 & 1.08& 3.49 & 0.53  \\
 20091117 & 55152.1 & 15.489 & 0.007 &  15.04 & 0.007 & 5.08 & 0.56 & 5.31 & 1.80 & 3.95 & 0.88  \\
  20091130 & 55165.0 & 14.847 & 0.007 &  14.372 & 0.006 & 4.12 & 0.88& 6.25 & 0.97 & 3.49& 0.37  \\
  20100519$ ^\dag$& 55335.4 & 15.394 & 0.02 &  15.032 & 0.007 & 4.43 & 0.66& 6.99 & 2.59 & 6.70 & 3.00   \\
  20100702$^\dag$ & 55379.4 & 15.715 & 0.012 &  15.367 & 0.011 & 6.87 & 1.09 & 7.39 & 2.72 & 9.84 & 4.28   \\
  20100714 & 55391.4 & 15.957 & 0.01 &  15.581 & 0.009 & 6.50 & 0.92 & 6.33 & 2.74 & 13.01 & 4.43  \\
  20100802$^\dag$ & 55410.4 & 15.88 & 0.011 & 15.378 & 0.01  & - & - & 10.57 & 5.91 & 9.27 & 6.67  \\
  20100920 & 55459.1 & 15.229 & 0.01 &  14.759 & 0.008 & 3.41 & 0.74& 4.97 & 1.22 & 1.89 & 0.38  \\
  20101112 & 55512.0& 14.427 & 0.007 &  13.937 & 0.006 & 2.04 & 0.29& 3.72 & 0.71 & 1.70 & 0.36  \\
  20101118 & 55518.1 & 14.065 & 0.009 &  13.519 & 0.005 & 1.89 & 0.46& 4.00 & 0.53 & 1.77 & 0.18  \\
  20101203 & 55533.0 & 15.282 & 0.019 &  14.682 & 0.007 & 3.99 & 0.82& 5.66 & 2.50 & 3.28 & 1.93   \\
  20110611 & 55723.4 & 16.295 & 0.017 &  15.917 & 0.016 & 11.20 & 2.84 & 11.24 & 6.38 & 21.70 & 10.94  \\
  20110626 & 55738.4 & 16.319 & 0.021 & 16.027 & 0.026 & 6.71 & 2.70 & 12.68 & 6.91 & 14.29 & 7.70  \\
 20111002 & 55836.2 & 16.529 & 0.02 & 16.206 & 0.021 & - & - & 15.97 & 6.35 & 20.74 & 8.61 \\
  20111017 & 55851.1 & 16.463 & 0.017 &  16.182 & 0.017 & 11.92 & 1.84 & 11.56 & 4.39 & 19.50 & 6.40   \\
  20111102 & 55867.1 & 16.604 & 0.023 &  16.208 & 0.021 & 12.56 & 1.55 & 16.65 & 5.34 & 26.88 & 7.91   \\
  20111202 & 55897.0 & 16.635 & 0.035 &  16.332 & 0.038 &  15.54 & 1.99 & 20.85 & 6.97& 17.00 & 6.04  \\
\enddata

\tablecomments{Columns give the (1) UTC date of the spectroscopy observation, (2) corresponding Modified Julian date (MJD),
(3) V magnitude (Vega), (4) 1$\sigma$ uncertainty in the V-band magnitude, (5) R magnitude, (6) 1$\sigma$ uncertainty in the R-band magnitude, (7, 9, 11) equivalent widths, and (8, 10, 12) 1$\sigma$ uncertainties of \mgii,\ H\bet\, and H\g, respectively. V- and R-band magnitudes were obtained from the \smarts\ 1.3m telescope+ANDICAM instrument. Spectra were taken with the \smarts\ 1.5m telescope + RCSpectrograph. All observations are contemporaneous, as the maximum amount of time between a given photometry and spectroscopy observation is never greater than two nights (maximum separation on 20100702); all other observations are within fractions of a day. $^\dag$\smarts\ photometry taken on 20091001 (55105),  20100519 (55335) 20100630 (55377), 20100803 (55411), respectively.\label{tab:dates}}
\end{deluxetable*}

\section{Observations and Data}
\label{sec:obs}

\subsection{Photometry \& Polarimetry}\label{sec:phot}
Optical photometric data were obtained using the Small and Medium Aperture Research Telescope System\footnotemark[1] (\smarts) 1.3m+ANDICAM in Cerro Tololo, Chile. The \smarts\ photometry and data analysis are described by \citet{Bonning12}. We use the same procedures here to derive the multi-band photometry of \thesource\ across the full 3.3-year period; these data are listed in Table~\ref{tab:dates}. 

We also use optical photometry and polarimetry obtained with the Steward Observatory\footnotemark[2]; the data analysis is described by \citep{Smith09}. The optical photometry represents the sum of emission from the presumably unpolarized accretion disk, and the more highly polarized jet emission. In the \smarts\ data, we were able to separate the two components using variability \citet{Bonning12}. Optical polarization is also useful in determining the synchrotron flux contribution to the optical continuum. We obtained the polarized flux by multiplying the percentage polarization by the optical V-band flux \citep[e.g.,][]{Smith94, Raiteri12}.

Figure~\ref{fig:lc} shows the light curves of \thesource\  from 2008 - 2011, from \fermi\ (E $>$ 100MeV) \g-ray, polarized optical flux (V) measured at Steward Observatory, and \smarts\ B- and J- band photometry. Vertical lines indicate dates when \smarts\ optical spectra were obtained. 
\footnotetext[1]{http://www.astro.yale.edu/smarts/glast/index.html}
\footnotetext[2]{http://james.as.arizona.edu/~psmith/Fermi/}

\begin{figure*}
\plotone{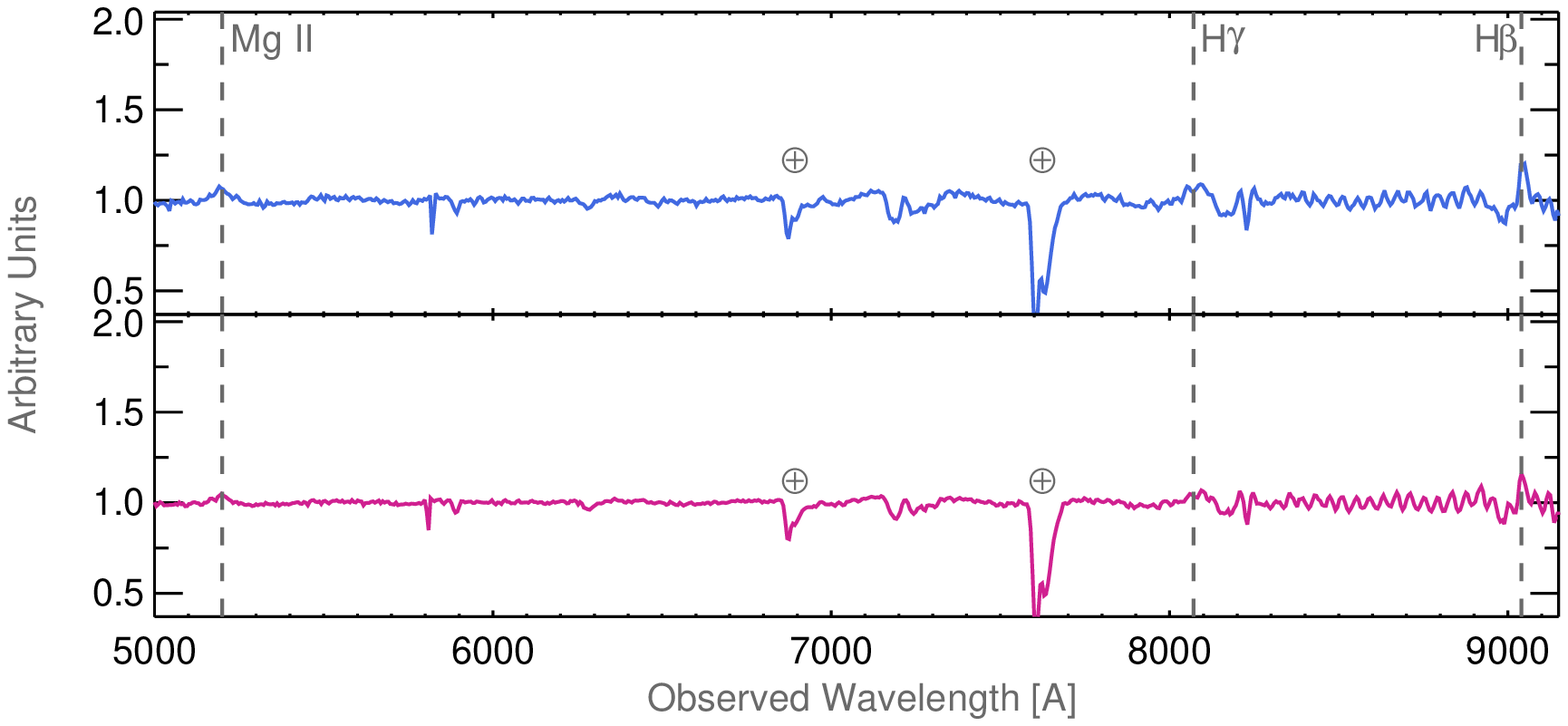} 
\figcaption{Two examples of \smarts\ optical spectra in the observed frame, obtained on 30 November 2009 (V = 14.85; \textit{top}) and 18 November 2010 (V = 14.07; \textit{bottom}). These spectra correspond to the exceptional \g-ray flares (log F$_{100}$ = -5.1 $\pm$ 0.03 and -4.2 $\pm$ 0.01 ph cm$^{-2}$s$^{-1}$, respectively) and emission line fluxes studied in detail in this work. Vertical lines show the (redshifted) wavelengths of identified emission lines. Spectra are normalized for presentation purposes only and telluric lines ($\earth$) are identified.  \label{fig:spectra}}
\end{figure*}

\subsection{\smarts\ Spectroscopy}\label{sec:spec}
Low resolution optical spectra (R $\sim$ 500) of \thesource\ (z=0.859) were obtained from 2008 August 22 through 2011 December 1, approximately twice monthly, as detailed in Table \ref{tab:dates}, using the \smarts\ Consortium 1.5m + RCSpectrograph in queue-scheduled mode. 
The 1.5-meter Cassegrain spectrograph is at an \textit{f}/7.5 focus with plate scale 18.1$^{\prime\prime}$ mm$^{-1}$ and LORAL 1K (1200 $\times$ 800) CCD. 
The primary grating for this study has first order resolution of 17.2~\ang, spectral coverage of 3600 - 9000 \ang, and slit width of 2$^{\prime\prime}$.  
In 2008-2010, the integration time was set to 900~s, with three sequential exposures per night. Beginning in 2011, exposure times were reduced to 600~s to allow observation of more sources per observing night. 
Two illustrative examples of \smarts\  optical spectra, during the two highest \g-ray flux states, can be seen in Figure \ref{fig:spectra}.

The spectra were bias- and overscan-subtracted then flat-fielded using standard \textit{IRAF} procedures. 
Wavelength calibration was completed using reference HeAr spectra taken immediately prior to the target integration. 
The spectra were then interpolated to a linear wavelength scale.
Bad pixels were rejected, then spectra were interpolated across them. The three nightly exposures were then averaged together to produce the highest possible signal-to-noise spectrum per night. 

The spectra were calibrated using near-simultaneous photometry from \smarts\ (see Table~\ref{tab:dates}). The continuum flux near the \mgii\ line ($\lambda_{obs}$ = 5200~\ang) was calibrated using the same \smarts\ V-band photometry (effective wavelength 5500~\ang) since the observed wavelength is redshifted into this band. 
Similarly, the Balmer lines were calibrated with the R-band photometry (H\g, $\lambda_{obs}$ = 8071~\ang; H\bet, $\lambda_{obs}$ = 9040~\ang; see Figure~\ref{fig:spectra}).
The conversion from each waveband magnitude to flux density, in units of erg s$^{-1}$cm$^{-2}$\ang$^{-1}$, utilized \citet{Bessell98} zero points. 
The equivalent width can be converted to an emission line flux by multiplying the flux density.
This conversion to line flux is the largest error introduced in our calculation.
We note that on 2010 May 20, although Steward had a more simultaneous photometric observation, we used the \smarts\ photometry due to calibration mismatches between the two datasets, which caused the R-band magnitudes to differ by more than the stated uncertainties.   

No order-blocking filters were used to obtain the spectra presented here. Given the width of the spectral range covered, second order contamination may contribute to the continuum flux at redder wavelengths greater than $\approx$ 6400 $\ang$, which could affect line flux measurements of the Balmer lines. An estimation of the second order contamination for this optical setup was obtained using historical observations of standard star LTT 4364 with and without the order blocking filter GG495 on 2004 April 20. We find the second order contamination to be $\approx$ 5\% at 8070  $\ang$ and $\approx$ 8\%  at 9040$\ang$, the observed locations of the Balmer lines in this study.  These estimates are in agreement with previous characterizations of second order contamination found in the literature \citep{Szokoly04, Stanishev07}. The second order contamination measured in the standard star is an upper limit on the contamination expected in \thesource\ given that it is often redder than the standard star. The measured equivalent width would change by, at most, a few percent which contributes less than other uncertainties in the line flux measurement. Thus second order contamination is not considered in the following analysis.

The line equivalent widths were measured by fitting a Gaussian to the emission lines above the continuum, minimizing the $\chi^2$ statistic, using the MPFIT package \citep{Markwardt09}. 
The continuum range used on each side of the line was 50 \ang\ wide.
The uncertainty in the equivalent width was determined by running 500 Monte Carlo simulations of the fitted line, including the measured noise in the count rate of each pixel. 
For each Monte Carlo simulation, the emission lines were fitted and equivalent widths were calculated. 
The reported error of the equivalent width of each line is the standard error on the 500 measured equivalent widths from the simulation. The low-resolution spectroscopy presented here does not allow the measurement of accurate line shape characteristics.

\begin{deluxetable}{ccc}\footnotesize
\tablecolumns{3}
\tablewidth{0pt}
\tablecaption{\g-ray Flare Periods for \thesource}
\tablehead{
\colhead{UT Date} &
 \colhead{MJD} &
\colhead{$\Delta$t (days)}  
}
\startdata
11 Sep 2008  - 11 Oct 2008 & 54720 - 54750 & 30 \\
07 Aug 2009 - 26 Sep 2009 & 55050 - 55100 & 50\\
15 Nov 2009 - 04 Jan 2010 & 55150 - 55200 & 50\\
20 Mar 2010 - 03 Jun 2010 & 55275 - 55350 & 125\\
31 Oct 2010 - 20 Dec 2010 & 55500 - 55550 & 50
\enddata
\tablecomments{$\Delta$t is the approximate duration (FWZM) of the \g-ray flare, in days. These flare windows are referenced in the subsequent line flux analysis. \label{tab:gam}}
\end{deluxetable}

\subsection{\fermi\ Observations}
\fermi/LAT data were obtained from the \fermi\ Science Support Center website\footnotemark[3] for 2008 August 04 - 2011 December 05. 
Pass 7 data (event class 2) were analyzed using \fermi\ Science Tools (v9r27p1) with user-contributed `LAT Analysis Scripts', which automate the reduction and likelihood analysis of the source.
 Galactic response functions (gal\_2yearp7v6\_0), isotropic diffuse background (iso\_p7v6source) and instrument response functions (P7SOURCE\_V6) were utilized in the analysis. 
Data were constrained to time periods where the zenith angle was less than 100$^\circ$ to avoid Earth limb contamination, and photons to within a 20$^\circ$ region centered on the source of interest.  
The \g-ray spectra of \thesource\ were modeled as a log parabola, with the photon flux and spectral index, $\alpha$, as free parameters; spectral curvature, \bet, remained a fixed parameter.
 \fermi\ light curves (E $>$ 100MeV) were calculated in one day time intervals, to match the average \smarts\ observation cadence. \footnotetext[3]{http://fermi.gsfc.nasa.gov/ssc/data/access}

Figure \ref{fig:lc} shows the complete light curve obtained from the present analysis, with the five flaring periods analyzed in this study identified in Table~\ref{tab:gam}.
Daily points for which TS $>$ 25 are plotted, where TS is the Fermi test statistic and is roughly equivalent to 5$\sigma$ detection level \citep{Mattox96,Abdo09}. Upper limits were obtained on the remaining dates shown.

\begin{figure}
\epsscale{0.95}
\plotone{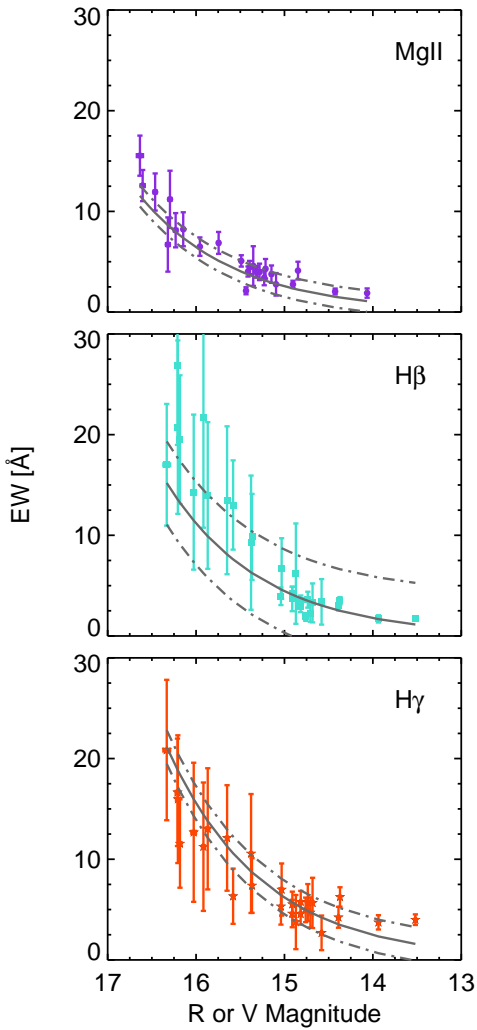} 
\figcaption{Emission line equivalent width versus continuum magnitude for \mgii, H\bet, and H\g, overlaid with best-fit constant emission line flux (solid line) and the 1$\sigma$ uncertainties (dot-dashed line). V-band continuum magnitude is used for \mgii\ and R-band continuum magnitude is used for the Balmer lines. The best-fit line is obtained by $\chi^2$ minimization, corresponding to constant line fluxes of 9.3$\times$10$^{-15}$ erg s$^{-1}$ cm$^{-2}$, 1.4$\times$~10$^{-14}$ erg s$^{-1}$ cm$^{-2}$, and 9.9$\times$~10$^{-15}$ erg s$^{-1}$ cm$^{-2}$ for \mgii, H\bet, and H\g, respectively. \label{fig:ew}}
\end{figure}

\begin{figure*}
\centering
\epsscale{1.3}
\plotone{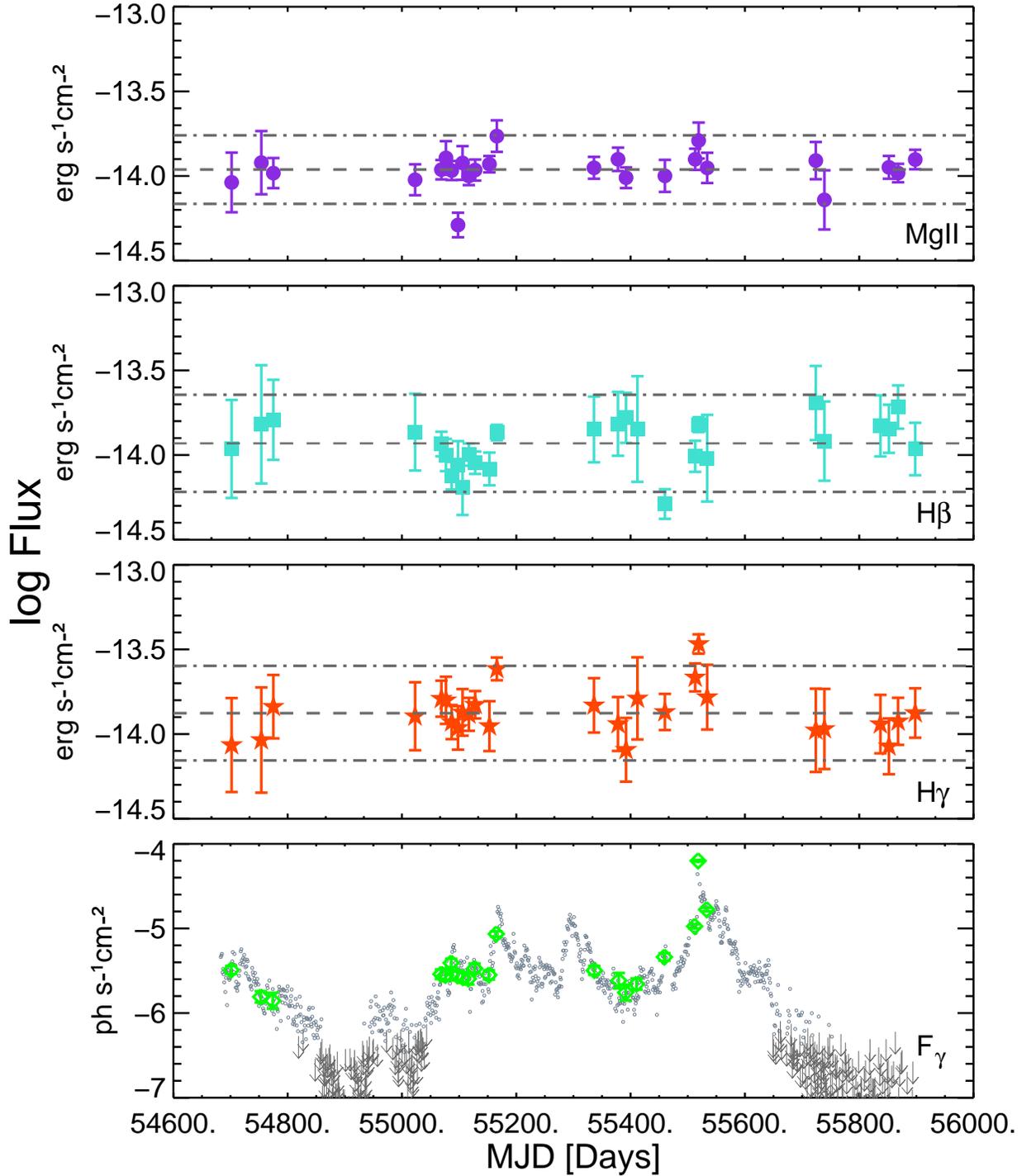}
\figcaption{Broad emission line flux light curves for \mgii\ (purple circles), H\bet\ (cyan squares) and H\g\ (orange stars). The bottom panel shows the \fermi\ \g-ray light curve (TS $>$25) for the same MJD (green diamonds) and over the total observed interval (grey points). The average flux of each emission line is represented by the dashed lines and 2$\sigma$ deviations are marked by dot-dashed lines. Over the 3.3 years of observation, the line fluxes deviate by more than 2$\sigma$ above the mean only on MJD 55165 and 55518 in \mgii\ and H\g, respectively. This lack of strong detectable variability in the line emission is in stark contrast to the factor of nearly 100 variations in \g-ray flux over the same time period, as seen in the bottom panel. However, the highest \g-ray flare phases (MJD 55167 and 55520) correspond to the greatest deviation in the H\g\ and \mgii\ line fluxes. The rise and fall of the H\g\ line flux, in particular, appears to trace the rise and fall of the \g-ray flux. 
 \label{fig:linelc}}
\end{figure*}

\begin{figure*}
\centering
\begin{tabular}{cc}
\epsscale{1.2}
\plotone{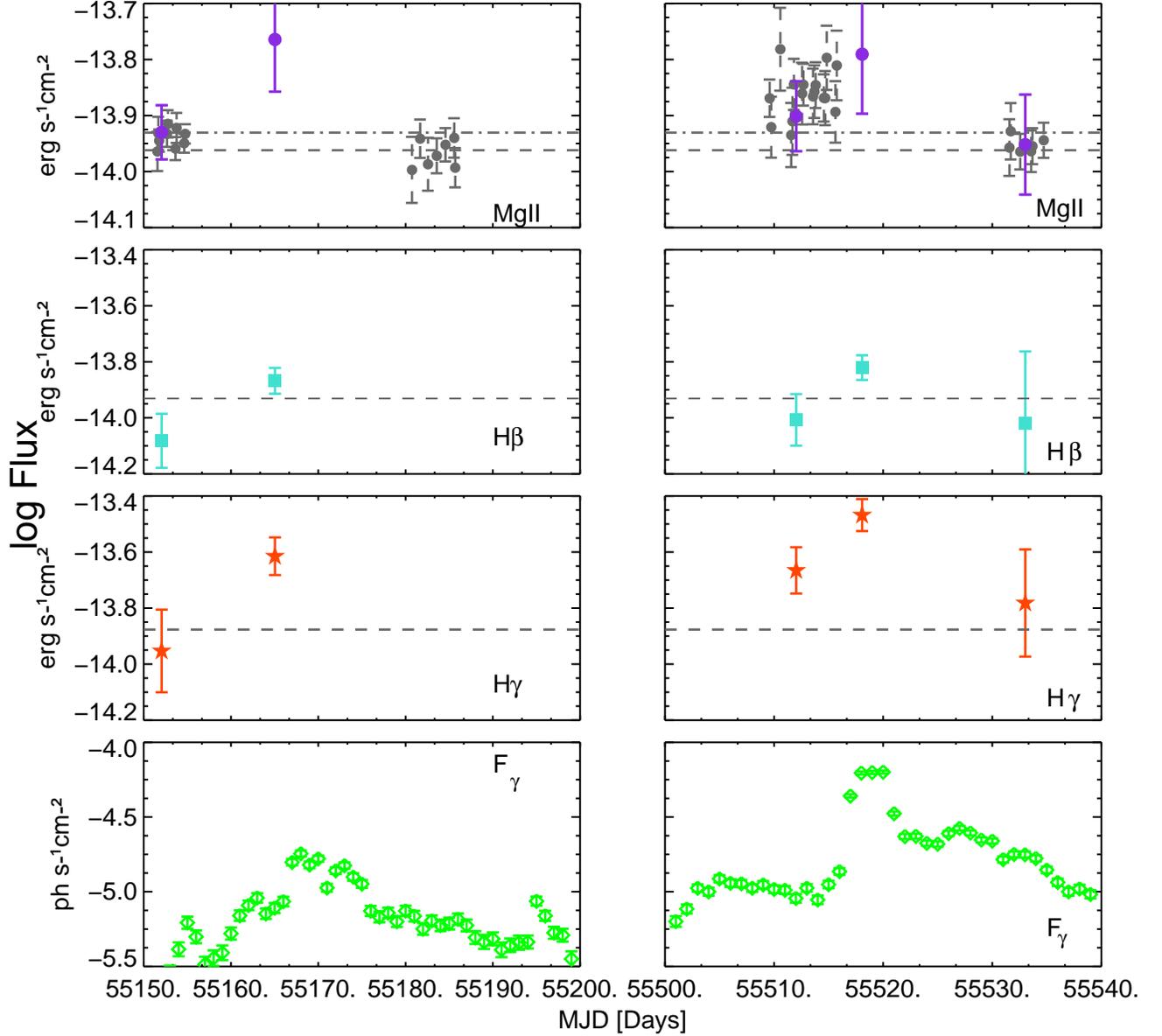}
\end{tabular}
\figcaption{Multiwavelength light curves during the 2009 December flaring period (MJD 55150 - 55200; \textit{left}) and 2010 November (MJD 55500 - 55540; \textit{right}). Panels are as in Figure \ref{fig:linelc}. Grey points are data from \citet{LT13} during the pre- and post-flare phases. The dashed error bars represent the reported errors in addition to the estimated error on the flux. Dashed and dot-dashed lines show the mean line flux values for the \smarts\ and \citet{LT13} data, respectively. The purple circles, cyan squares, and orange stars show the \smarts\ data obtained during the flaring phase, which collectively show a significant deviation from the mean in the \mgii\ and H\g\ line flux in the same sense as the \g-rays (green diamonds). This suggests that the jet continuum, in its brightest state, contributes significantly to the photoionization of the broad line gas. The \citet{LT13} data were scaled on a quiescent date to match the \smarts\ data, given that the \smarts\ \mgii\ line flux was not disentangled from the FeII emission and thus is systematically higher, as well as to minimize the effect of unknown flux calibration errors in those light curves. \label{fig:F3linelc} The \smarts\ data show significant deviations in both the \mgii\ and H\g\ line flux, with H\g\ showing a factor of 2.5 increase in flux during the flaring phase.}
\end{figure*}

\begin{figure*}
\epsscale{1.2}
\plotone{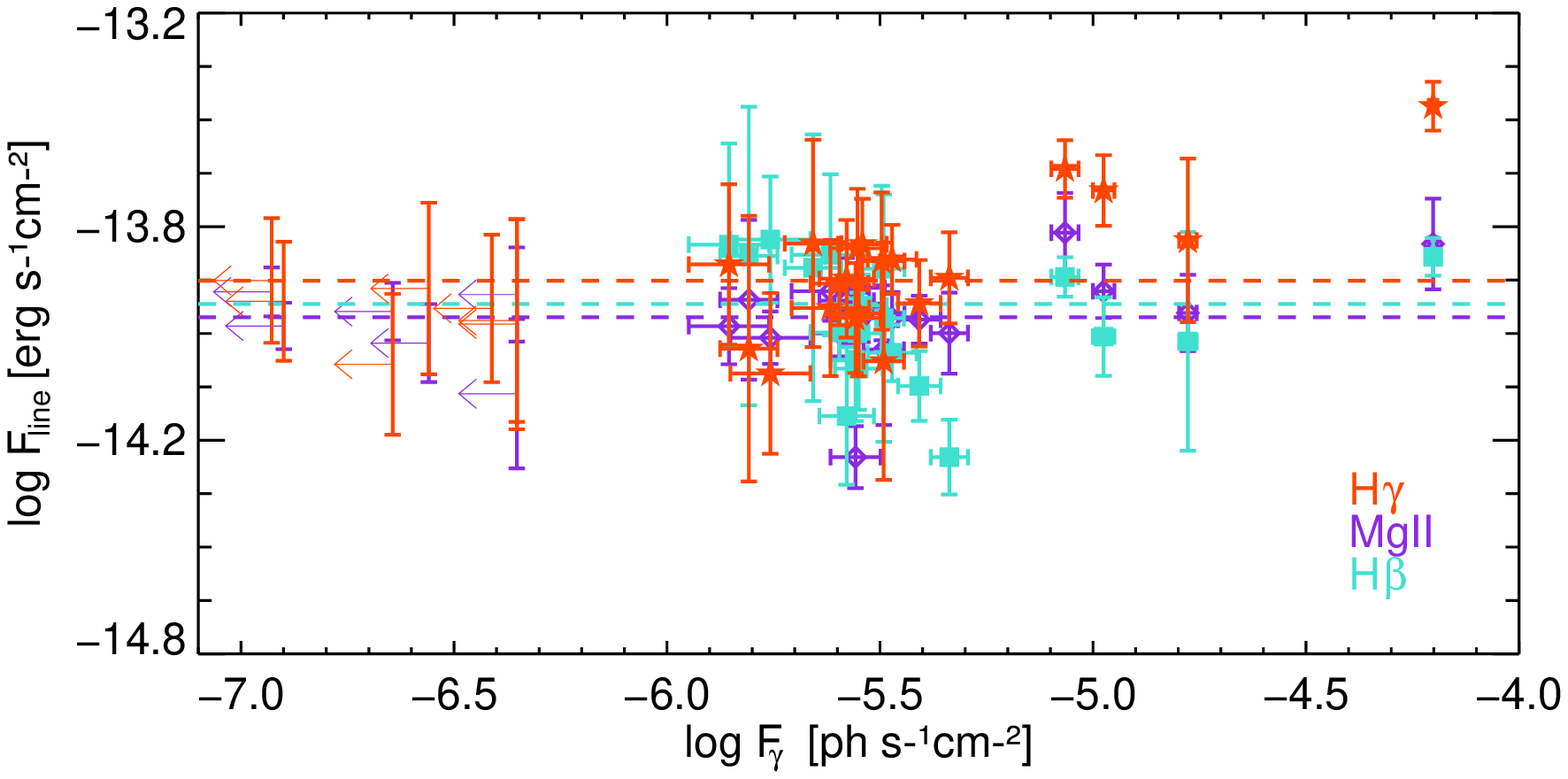}
\figcaption{Broad emission line fluxes vs \g-ray flux for \mgii\ (\textit{purple circles}), H\bet\ (\textit{cyan squares}), and H\g\ (\textit{orange stars}).  Arrows mark where the \fermi\ confidence level, TS is $<$ 25 (and hence upper limits in \g-ray flux); points are dates for which TS $>$ 25. The average flux for each emission line is shown with the dashed line with corresponding colors. The places of highest detectable deviation from the mean line flux values are coincident with the very highest \g-ray fluxes, perhaps indicating a jet contribution to the photoionization of the broad line gas, but the emission line flux is remarkably constant across 3 orders of magnitude in \g-ray flux, suggesting that the jet is not a significant source of photoionization most of the time. \label{fig:gamvlines}}
\end{figure*}

\section{Results and Analysis}
\label{sec:results}

\subsection{Emission Line Variability}\label{sec:elv}
Figure \ref{fig:ew} shows the equivalent width versus optical magnitude for the \mgii\ (purple circles), H\bet\ (cyan squares), and H\g\ (orange stars) emission lines over the total observation period. 
The data were compared to models of constant line flux in each emission line.
The best-fit constant line flux is found by minimizing the $\chi^2$ statistic for a family of constant emission line fluxes ranging from 7$\times$10$^{-15}$ erg s$^{-1}$cm$^{-2}$ to 20$\times$10$^{-15}$ erg s$^{-1}$cm$^{-2}$ in 0.1$\times$10$^{-15}$ increment steps.  
Best-fit constant lines of 9.3$\times$10$^{-15}$ erg s$^{-1}$ cm$^{-2}$, 1.4$\times$~10$^{-14}$ erg s$^{-1}$ cm$^{-2}$, and 9.9$\times$~10$^{-15}$ erg s$^{-1}$ cm$^{-2}$ were found for \mgii, H\bet, and H\g, respectively. 
While the continuum flux increases, the emission line fluxes show little significant deviation from the constant line flux approximation. 

Emission line light curves for \mgii, H\bet, and H\g\ are shown in Figure \ref{fig:linelc}, compared to the \fermi\ \g-ray light curve simultaneous to within one day. Dashed lines are mean flux values and dot-dashed lines are the $\pm$2$\sigma$ uncertainties. 
From 2008 August through 2011 December, the emission line fluxes are largely consistent with constant line flux, except during two epochs of observation, shown in Figure \ref{fig:F3linelc}.

Figure \ref{fig:F3linelc} (\textit{left}) shows the 2009 December \g-ray flaring period (MJD 55150 - 55200). During the 2009 December flare (peak \g-ray flare at MJD 55167), we see positive spectral deviations of 1.2$\sigma$ in H\bet, 1.8$\sigma$ in the \mgii\ line and 2.8$\sigma$ in the H\g\ line (at MJD 55165), corresponding to a factor of 1.7 from the mean flux for both \mgii\ and H\g. Grey points in the line flux plots mark data obtained from Steward Observatory, as reported by \citet{LT13}, which extend from 4000 - 7550 \ang\ rest frame waveband coverage, so that the \mgii\ line is the only emission line that can be compared with our findings. The \citet{LT13} data appear to have significantly smaller error bars than we report here, which may be due to the fact that the flux calibration errors are not included in their published data tables. We estimate the additional error due to flux calibration in their data, in addition to the reported error, and plot these uncertainties with the light grey dotted error bars. H\g\ and H\bet\ are reported only in the current paper. While no significant deviation in the \mgii\ line is reported in the \citet{LT13} data for this flaring period, we note that those observations occured immediately pre-flare (MJD 55155) and so do not characterize the response of the spectrum during the peak flaring period itself.

Figure \ref{fig:F3linelc} (\textit{right}) expands the 2010 November flaring period (MJD 55500 - 55540) when \thesource\ was the brightest source in the \g-ray sky. 
\citet{LT13} reported a 40\% increase in the \mgii\ line flux, which we see as a 1.5$\sigma$ deviation from the mean. 
In addition, we see a factor of 2.5 increase in the line flux of H\g\ (corresponding to a 3.7$\sigma$ deviation) during the same epoch (MJD 55518). 
We note that while the increase in \mgii\ line flux is similar to that in the 2009 December flare, H\g\ showed a more significant increase in the line flux during the more powerful \g-ray flaring period in 2010 November.

During both flaring periods, the measured continuum flux was measured to be significantly higher than would be predicted if the line fluxes were assumed to be constant, indicating that the measured line fluxes were larger than the mean line flux in each case. For the 2009 December flaring period, the predicted continuum flux differed from the empirical line flux at the peak line flux by 69\% (Mg II), 77\% (H$\beta$), and 56\% (H$\gamma$), corresponding to an average of 4$\sigma$ difference; similarly for the 2010 November flaring period the deviations were 32\% (Mg II), 65\% (H$\beta$), and 15\% (H$\gamma$), an average of 2$\sigma$ difference between the measured and predicted continuum flux. Thus, the line flux variations are distinguishable from increases in continuum flux alone, as can be seen in Figure~\ref{fig:ew}. We determine the errors on the predicted line flux by summing the uncertainties in the equivalent width and photometry in quadrature (see Table~\ref{tab:dates}). 

Figure \ref{fig:gamvlines} shows emission line flux versus the \g-ray flux; there is clear evidence that high jet flux states contribute to the photoionizing flux, although the jet cannot contribute the majority of the photoioinizing flux. Otherwise, the quiescent jet states would correspond to decreased line flux, which is not seen. 
Instead, the data remain near the mean emission line flux in all three emission lines during quiescent jet states.

\subsection{Broad Line Region Variability}\label{sec:blr}
The three emission lines discussed in this work can be combined to estimate the total broad line and accretion disk luminosities.
The total broad line region luminosity, \lblr, can be estimated empirically by extrapolation from a few individual lines. 
The six strongest broad emission lines in quasars ---
Ly$\alpha$ (100), CIV (63), \mgii\  (34), H\g(+OIII) (13), H\bet\  (22), and H$\alpha$ (77) 
--- represent $\sim$ 60\% of the total luminosity emitted in broad lines \citep{Celotti97}, 
where the parentheses indicate the observed line strengths relative to Ly$\alpha$ \citep{Francis91, Gaskell81}.
Using these line ratios, and summing the three lines that are detected in virtually all spectra obtained for this study
--- \mgii, H\g, and H\bet\ --- 
we estimate the total broad-line luminosity: 
\begin{equation}
L_{BLR} \simeq \eta (L_{MgII} + L_{H\beta} + L_{H\gamma}) .
\end{equation}
where $\eta$ = 7.5 when all three lines are used, and 14.7 when only the Balmer lines are detected. 
The line luminosities obtained for each spectrum are listed in Table~\ref{tab:lums}, and the mean broad line luminosity is \lblr\ $\simeq$ 10$^{45}$ erg s$^{-1}$ (see Table~\ref{tab:dqtys}). 
Using three lines makes the broad line luminosity estimate more robust than the usual correction from one line \citep{Celotti97,Sbarrato12}.
We note that the line ratios suggested by \citet{Celotti97} are consistent for the three lines we observed in the \smarts\ spectra.

\begin{deluxetable}{lcccc}
\tablewidth{0pt}
\tablecaption{Emission Line Luminosities$^1$}
\tablehead{\colhead{UTC} & \colhead{MJD} & \colhead{log L$_\mgii$} & \colhead{log L$_{H\beta}$} &  \colhead{log L$_{H\gamma}$} } 

\startdata
20080823& 54701.2 & 43.5 & 43.6  & 43.4  \\
20081014 & 54753.1 & 43.6 & 43.8  & 43.5    \\
20081104 & 54774.0 & 43.6 & 43.8  & 43.7  \\
20090710 & 55022.4 & 43.5 & 43.7 & 43.7 \\
20090825 & 55068.2 & 43.6&  43.6 & 43.8   \\
20090902 & 55076.2 & 43.7 &  43.6  & 43.8    \\
20090912 & 55086.1 & 43.6 & 43.4  & 43.6   \\
20090923 & 55097.1 & 43.3 &  43.5 &  43.6  \\
20091002 & 55106.2 & 43.7 &  43.4 &  43.8   \\
20091012 & 55116.1 & 43.6 &  43.6 &  43.7   \\
20091023 & 55127.1 & 43.6 &  43.5  & 43.7   \\
20091117 & 55152.1 & 43.6 &  43.5 &  43.6   \\
20091130 & 55165.0 & 43.8 &  43.7  & 43.9   \\
20100520 & 55336.4 & 43.6 & 43.7 & 43.7  \\
20100702 & 55379.4 & 43.6 & 43.7  & 43.6\\
20100714 & 55391.4 & 43.6 & 43.8 &  43.5  \\
20100802 & 55410.4 & - & 43.7  & 43.8  \\
20100920 & 55459.1 & 43.5 &  43.3 &  43.7  \\
20101112 & 55512.0 & 43.7  & 43.6  & 43.9  \\
20101118 & 55518.1 & 43.8  & 43.7 & 44.1  \\
20101203 & 55533.0 & 43.6 &  43.5  & 43.8   \\
20110611 & 55723.4 & 43.7 &  43.9 & 43.6\\
20110626 & 55738.4 & 43.4 &  43.6 & 43.6  \\
20111002 & 55836.2 & - & 43.7 & 43.6 \\
20111017 & 55851.1 & 43.6 &  43.7 &  43.5   \\
20111102 & 55867.1 & 43.6 & 43.8 &  43.6   \\
20111202 & 55897.0 & 43.7 & 43.6 &  43.7   \\
\enddata

\tablecomments{The UTC is in YYYYMMDD format and all luminosities in units of erg s$^{-1}$. \label{tab:lums}}
\end{deluxetable}

\begin{deluxetable}{ccccc}\footnotesize
\tablecolumns{3}
\tablewidth{0pt}
\tablecaption{Derived Physical Quantities.}
\tablehead{ 
 \colhead{Parameter} &
 \colhead{} &
 \colhead{} &
 \colhead{Units}}  
\startdata
$\langle$F$_\mgii\rangle$ & -13.96 (0.10) & & erg s$^{-1}$ cm$^{-2}$\\
$\langle$F$_{H\beta}\rangle$& -13.93 (0.14) & & erg s$^{-1}$ cm$^{-2}$ \\
$\langle$F$_{H\gamma}\rangle$& -13.87 (0.14) & & erg s$^{-1}$ cm$^{-2}$\\
$\langle$\lblr$\rangle$ & 45 &  &erg s$^{-1}$\\
\enddata
\tablecomments{The mean line flux and broad line luminosity derived for \thesource. All quantities are in rest frame units. Average fluxes and luminosities are given in logarithmic units. \label{tab:dqtys}}
\end{deluxetable}

\section{Discussion}
\label{sec:disc}
Our results show that the broad line region in \thesource, as traced by the \mgii, H\bet, and H\g\ emission lines, is not undergoing large variations over 90\% of the period of observation. 
This is consistent with a slowly varying accretion disk providing a large fraction of the photoionizing flux in the broad line region. 
If the accretion disk were undergoing significant, changes in accretion rate, the (disk) photoionizing flux would change, and in turn the emission line fluxes would also vary, likely with lags on the order of several months. Given the lack of large amplitude variability in the broad lines and thus in the photoionizing disk emission, the jet does not contribute much photoionizing flux to the broad line region in general. This is consistent with the 3C279 result \citep{Koratkar98}, showing only small line flux variations ($\approx$25\%) with factors of 50 in optical continuum variability. 

Historically, the UV thermal bump from disk emission has been observed to be $\sim$ 7 $\times$10$^{46}$ erg s$^{-1}$ in the faint state of \thesource\ following the 2005-2006 outburst \citep{Villata07, Raiteri07}. Comparing our estimate of the total broad-line luminosity, $\approx$1$\times$10$^{45}$erg s$^{-1}$, we infer that either the covering factor is $\approx$ 1/70 or, in order for the covering factor to be closer to the 0.1 observed in other quasars \citep{Baldwin78, Smith88}, the disk luminosity must have declined by a factor of 7. The lack of variability seen in the UV big blue bump during both active and quiescent jet states \citep{Bonnoli11, Bonning12}, however, favors the former explanation.

However, the very highest \g-ray jet flaring periods are associated with significant increases in the broad line region emission. 
The 2009 December flare phase (MJD 55166 - 55173), F$_{100}$ = 22$\times$10$^{-6}$ ph cm$^{-2}$s$^{-1}$ \citep{Ackermann10}, corresponded to significant increases in the \mgii\ and H\g\ line fluxes during the plateau phase of that flare period. The 40\% increase in \mgii\ line flux seen by \citet{LT13} in 2010 November occurs during the plateau phase (MJD 55503 - 55515), whereas we saw a larger increase. 
The 2$\sigma$ deviation in \mgii\ and the 4$\sigma$ increase in H\g\ during the flaring phase (MJD 55516 - 55522, F$_{100}$ = 66$\times$10$^{-6}$ ph cm$^{-2}$s$^{-1}$) are much larger than the increase seen by  \citet{LT13}; these results are consistent because they do not sample the same time periods, and in particular, they do not sample the \g-ray flaring period (as defined by the Ackermann-Abdo convention\footnotemark[4]). \footnotetext[4]{The 5 major flaring periods in \thesource\ show a consistent phenomenological \g-ray structure that includes pre-flare, plateau, flare and post-flare phases evident in the 2009 December, 2010 April and 2010 November flaring periods \citet{Ackermann10,Abdo11}.} In our case, because we normalize using highly accurate photometry, we avoid systematic errors that come from the difficulty of absolutely calibrating the spectra, and maintaining calibration accuracy over long time intervals.

The 2010 April flare (MJD 55280 - 55300, F$_{100}$ = 16$\times$10$^{-6}$ ph cm$^{-2}$s$^{-1}$) showed no significant line variability in any of the lines analyzed in this work.
While the \fermi\ \g-ray fluxes for the 2009 December and 2010 April flaring phases are similar to the 2010 November plateau phase intensity, the 2010 April flare has slightly lower peak intensity and is the only flare that did not show a detectable 7mm core ejection event \citep{Jorstad12}. 
Furthermore, in contrast to the significant increases in polarization during the 2009 and 2010 flares, the 2010 April flare also lacks significant increase in the polarized (non-thermal) flux. This flare may therefore be produced in a different manner than the other two flares discussed here.
As such, the difference in line flux response may be due to lack of strong \g-emitting emission originating within the broad line region. The low polarization seen near MJD 55700, which also corresponds to low \fermi\ \g-ray flux also suggest no \g-emitting plasma is present. This is in contrast to the high polarization that is seen during high \g-ray states and when emission line variability is observed. 

The H\g\ line shows stronger variations than \mgii\ and H\bet\ in both the 2009 December and 2010 November flaring periods. 
This may be because the \mgii\ and H\bet\ lines are more challenging to measure due to systematic effects. 
H\bet\ suffers from a fall-off in detector sensitivity at the red end of the spectrograph; \mgii\ is diluted by the underlying Fe~II emission known to be present at the same wavelength interval (this systematic effect has not been accounted for in the quoted uncertainties of the \mgii\ line strength). 
The Fe~II contribution is expected to be strongest in the wings of the \mgii\ line and at a local minimum around the center of the \mgii\ line \citep{Wills85, Goad99}, and thus is not likely to contribute significantly to our line flux measurements.
In addition, \citet{LT13} decomposed the Fe II emission but it showed the same variation as the \mgii\ line (see their Figure 2), suggesting that its contribution does not obscure the emission line variation or else that it is not possible to disentangle it from the \mgii.
The lesser variability in \mgii\ compared to H\g\ is likely due to the stratification of the broad line region with respect to ionization parameter and gas density \citep{Clavel91, Goad99}. It should also be noted that H\g\ falls within the highly variable  atmospheric water vapor absorption feature from $\sim$8000 - 8400 $\ang$, which could in principle affect the line flux we measure. However, any atmostpheric absorption would attenuate the measured emission line flux, such that the true line flux is greater than that reported.

As the broad emission lines vary by less than a factor of 2 over most of the observation and the sampling is not dense enough, it is not possible to do reverberation mapping. 
However, constraints can be placed on the variability timescale during the 2010 November flare period given the coincidence of the \g-ray flare period with the deviation from the mean line fluxes. 
The peak \g-ray flux occurs within 2 days of the peak H\g\ line flux (MJD 55520 and 55518, respectively), so the light travel time from the broad line clouds to the jet acceleration region is $\le$ 2$\delta$ days. 
This work shows that the \mgii\ emission coincides with the strongest flaring period in 2010 November, as does the H\g\ emission and both lines increase near the peak of the 2009 December \g-ray flaring period. 
This suggests that during the strongest flaring periods the jet is bright enough to significantly increase the photoionizing flux in the broad line region, as also suggested by \citet{LT13}. If the disk were responsible for the changes we see, we would expect each flare to be associated with changes in the broad line region luminosity with a marked delay, which is not observed.

Previous studies have shown correlations among the millimeter, infrared, optical, ultraviolet and \g-ray emission \citep{Bonning09,Jorstad12, Wehrle12, Abdo11, Vercellone11} during the 2010 November flaring period.  
These correlations argue for the cospatiality of the synchrotron radiation with the lower energy photons on which inverse Compton scattering takes place.  That the variability of the broad line region luminosity coincides with the strongest \g-ray flaring periods suggests that the broad line region must lie further from the central source than the \g-emitting region, at least during the 2009 December and 2010 November flaring events. 
The fact that we see an increase in broad line strength at the same time as the Doppler boosted jet emission implies that broad-line clouds along the line-of-sight are affected by the jet flare. 

The location of the \g-emitting region is actively debated. Many studies suggest that it lies within the broad line region \citep{Bottcher07, Kataoka08, Tavecchio10, Ghisellini10, Poutanen10, Tavecchio10, Stern11, Abdo11}, which is at sub-pc scales (0.03 - 1 pc) \citep[e.g.]{Peterson93, Peterson06}, while others conclude it is well outside the broad line region, tens of pc from the black hole \citep{Marscher08, Jorstad10, Jorstad12, Agudo12}. We know from VLBI imaging that at least some jet flares are located at large distances from the central source ($\sim$10 pc) \citep[e.g., TEMZ model][]{Marscher11, Agudo11}. Following the 2009 December and 2010 November flaring periods, \citet{Tavecchio10, Abdo11} suggest the \g-emitting region of \thesource\ could be located at the outer edges of the broad line region (r$_{em} \sim$ 0.14 pc), using \g\g-opacity arguments. 

An independent argument can be used to determine whether the \g-emitting region could exist within the broad line region.
The correlation of the millimeter and \g-ray light curves imposes an upper limit on the synchrotron self-absorption frequency to be $\sim$ 7mm. 
The SED for \thesource\, presented in \citet{Bonning12} was fit with a one-zone synchrotron and inverse Compton model \citep{Coppi92, Ghisellini07} with parameters: size, R$_\gamma$ $\sim$ 10$^{17}$ cm; Doppler factor, $\delta\sim$20;  and magnetic field strength, B $\sim$ 1 G, similar to the fits of \citet{Bonnoli11}. For this model, the self-absorption frequency, $\nu_{ssa}$ $\le$ 3 mm, compatible with a \g-emitting region located within the canonical broad line region.

The beaming of the jet emission adds photoionizing flux to a small fraction of the gas in the broad line region, similar to that described in the mirror model \citep{Ghisellini96}. A systematic study of other blazars would confirm whether increased broad line fluxes are generally associated with the strongest jet flares. Then, by comparing the line variability characteristics of other (misaligned) quasars and radio galaxies, one could have an independent probe of the parent population of blazars.		

\section{SUMMARY AND CONCLUSIONS}
\label{sec:sum}
The optical-IR continuum and optical emission line flux variability of the blazar \thesource\ was measured throughout 3.3 years of continuous \fermi\ observations. 
Five dramatic \g-ray flares coincided with large increases in the mm, optical-IR, and ultraviolet bands but the broad emission lines varied much less, being consistent for the most part with constant line luminosity. Two notable exceptions were seen however, during the two strongest \g-ray flares. The coincidence of the increases in the broad emission lines at the peak of the two highest \g-ray flaring states (and core ejections) suggests that the \g-emitting region, at least in those flares, must be interior to the broad line region gas. 

To strengthen these results, much finer cadence optical spectroscopy is needed during the highest \g-ray flaring states of \thesource, meaning at least daily monitoring during and in the weeks immediately following a large \g-ray flare. Other bright blazars should be monitored spectroscopically to see whether they show similar evidence of photoionization by a compact gamma-ray source. We have undertaken such a study, with results to be reported in a future paper.\\

\acknowledgements
The authors wish to thank the anonymous referee for valuable improvements to this manuscript. We also wish to thank Frederick Walter, queue scheduler for the \smarts\ 1.5m+RCSpec, for providing the order blocking data and conversations regarding the performance of the spectrograph. \smarts\ observations of LAT-monitored blazars are supported by 
Yale University and \fermi\ GI grant NNX 12AP15G. J.C.I. has received support from NASA-Harriet Jenkins Pre-doctoral Fellowship Program, NSF Graduate Research Fellowship Program (DGE-0644492), and the National Research Council's Ford Foundation Dissertation Fellowship.  C.D.B., M.M.B, and the \smarts\ 1.3m and 1.5m observing queue also receive support from NSF grant AST-0707627. We are grateful for photometry and polarimetry from Paul Smith's monitoring program at the Steward Observatory, which is supported by Fermi Guest Investigator grants NNX08AW56G, NNX09AU10G, and NNX12AO93G.

\bibliography{refs2012}

\begin{thebibliography}{78}
\expandafter\ifx\csname natexlab\endcsname\relax\def\natexlab#1{#1}\fi

\bibitem[{{Abdo} {et~al.}(2009){Abdo}, {Ackermann}, {Ajello}, {Atwood},
  {Axelsson}, {Baldini}, {Ballet}, {Barbiellini}, {Bastieri}, {Baughman},
  {Bechtol}, {Bellazzini}, {Blandford}, {Bloom}, {Bonamente}, {Borgland},
  {Bouvier}, {Bregeon}, {Brez}, {Brigida}, {Bruel}, {Burnett}, {Caliandro},
  {Cameron}, {Caraveo}, {Casandjian}, {Cavazzuti}, {Cecchi}, {Charles},
  {Chekhtman}, {Chen}, {Cheung}, {Chiang}, {Ciprini}, {Claus}, {Cohen-Tanugi},
  {Colafrancesco}, {Collmar}, {Cominsky}, {Conrad}, {Costamante}, {Cutini},
  {Dermer}, {de Angelis}, {de Palma}, {Digel}, {do Couto e Silva}, {Drell},
  {Dubois}, {Dumora}, {Farnier}, {Favuzzi}, {Fegan}, {Ferrara}, {Finke},
  {Focke}, {Foschini}, {Frailis}, {Fuhrmann}, {Fukazawa}, {Funk}, {Fusco},
  {Gargano}, {Gasparrini}, {Gehrels}, {Germani}, {Giebels}, {Giglietto},
  {Giommi}, {Giordano}, {Giroletti}, {Glanzman}, {Godfrey}, {Grenier},
  {Grondin}, {Grove}, {Guillemot}, {Guiriec}, {Hanabata}, {Harding}, {Hartman},
  {Hayashida}, {Hays}, {Healey}, {Horan}, {Hughes}, {J{\'o}hannesson},
  {Johnson}, {Johnson}, {Johnson}, {Johnson}, {Kadler}, {Kamae}, {Katagiri},
  {Kataoka}, {Kerr}, {Kn{\"o}dlseder}, {Kocian}, {Kuehn}, {Kuss}, {Lande},
  {Latronico}, {Lemoine-Goumard}, {Longo}, {Loparco}, {Lott}, {Lovellette},
  {Lubrano}, {Madejski}, {Makeev}, {Massaro}, {Mazziotta}, {McConville},
  {McEnery}, {McGlynn}, {Meurer}, {Michelson}, {Mitthumsiri}, {Mizuno},
  {Moiseev}, {Monte}, {Monzani}, {Moretti}, {Morselli}, {Moskalenko}, {Murgia},
  {Nolan}, {Norris}, {Nuss}, {Ohsugi}, {Omodei}, {Orlando}, {Ormes}, {Ozaki},
  {Paneque}, {Panetta}, {Parent}, {Pelassa}, {Pepe}, {Pesce-Rollins}, {Piron},
  {Porter}, {Rain{\`o}}, {Rando}, {Razzano}, {Razzaque}, {Reimer}, {Reimer},
  {Reposeur}, {Reyes}, {Ritz}, {Rochester}, {Rodriguez}, {Romani}, {Ryde},
  {Sadrozinski}, {Sanchez}, {Sander}, {Saz Parkinson}, {Scargle}, {Schalk},
  {Sellerholm}, {Sgr{\`o}}, {Shaw}, {Smith}, {Smith}, {Spandre}, {Spinelli},
  {Starck}, {Strickman}, {Suson}, {Tajima}, {Takahashi}, {Takahashi}, {Tanaka},
  {Taylor}, {Thayer}, {Thayer}, {Thompson}, {Tibaldo}, {Torres}, {Tosti},
  {Tramacere}, {Uchiyama}, {Usher}, {Vilchez}, {Villata}, {Vitale}, {Waite},
  {Winer}, {Wood}, {Ylinen}, \& {Ziegler}}]{Abdo09}
{Abdo}, A.~A., {Ackermann}, M., {Ajello}, M., {et~al.} 2009, \apj, 700, 597

\bibitem[{{Abdo} {et~al.}(2011){Abdo}, {Ackermann}, {Ajello}, {Allafort},
  {Baldini}, {Ballet}, {Barbiellini}, {Bastieri}, {Bellazzini}, {Berenji},
  {Blandford}, {Bloom}, {Bonamente}, {Borgland}, {Bouvier}, {Bregeon},
  {Brigida}, {Bruel}, {Buehler}, {Buson}, {Caliandro}, {Cameron}, {Caraveo},
  {Casandjian}, {Cavazzuti}, {Cecchi}, {Charles}, {Chekhtman}, {Cheung},
  {Chiang}, {Ciprini}, {Claus}, {Conrad}, {Cutini}, {D'Ammando}, {de Angelis},
  {de Palma}, {Dermer}, {Digel}, {Silva}, {Drell}, {Dubois}, {Dumora},
  {Escande}, {Favuzzi}, {Fegan}, {Ferrara}, {Fortin}, {Fukazawa}, {Fusco},
  {Gargano}, {Gasparrini}, {Gehrels}, {Germani}, {Giglietto}, {Giommi},
  {Giordano}, {Giroletti}, {Glanzman}, {Godfrey}, {Grenier}, {Grove},
  {Guiriec}, {Hadasch}, {Hayashida}, {Hays}, {Horan}, {Itoh},
  {J{\'o}hannesson}, {Johnson}, {Kamae}, {Katagiri}, {Kataoka},
  {Kn{\"o}dlseder}, {Kuss}, {Lande}, {Larsson}, {Latronico}, {Lee}, {Longo},
  {Loparco}, {Lott}, {Lovellette}, {Lubrano}, {Madejski}, {Makeev},
  {Mazziotta}, {McConville}, {McEnery}, {Michelson}, {Mitthumsiri}, {Mizuno},
  {Moiseev}, {Monte}, {Monzani}, {Morselli}, {Moskalenko}, {Murgia},
  {Naumann-Godo}, {Nishino}, {Nolan}, {Norris}, {Nuss}, {Ohsugi}, {Okumura},
  {Orlando}, {Ormes}, {Paneque}, {Pelassa}, {Pesce-Rollins}, {Pierbattista},
  {Piron}, {Porter}, {Rain{\`o}}, {Rando}, {Razzaque}, {Reimer}, {Reimer},
  {Ritz}, {Roth}, {Sadrozinski}, {Sanchez}, {Scargle}, {Schalk}, {Sgr{\`o}},
  {Siskind}, {Smith}, {Spandre}, {Spinelli}, {Strickman}, {Takahashi},
  {Takahashi}, {Tanaka}, {Tanaka}, {Thayer}, {Thayer}, {Thompson}, {Tibaldo},
  {Torres}, {Tosti}, {Tramacere}, {Troja}, {Vandenbroucke}, {Vasileiou},
  {Vianello}, {Vilchez}, {Vitale}, {Waite}, {Wang}, {Winer}, {Wood}, {Yang}, \&
  {Ziegler}}]{Abdo11}
---. 2011, \apjl, 733, L26

\bibitem[{{Ackermann} {et~al.}(2010){Ackermann}, {Ajello}, {Baldini}, {Ballet},
  {Barbiellini}, {Bastieri}, {Bechtol}, {Bellazzini}, {Berenji}, {Blandford},
  {Bonamente}, {Borgland}, {Bregeon}, {Brigida}, \& {Bruel}}]{Ackermann10}
{Ackermann}, M., {Ajello}, M., {Baldini}, L., {et~al.} 2010, \apj, 721, 1383

\bibitem[{{Agudo} {et~al.}(2012){Agudo}, {Marscher}, {Jorstad}, \&
  {Gomez}}]{Agudo12}
{Agudo}, I., {Marscher}, A.~P., {Jorstad}, S.~G., \& {Gomez}, J.~L. 2012, in,
  Highlights of Spanish Astrophysics VII, Proceedings of the X Scientific
  Meeting of the Spanish Astronomical Society. (arXiv:1210.2234)

\bibitem[{{Agudo} {et~al.}(2011){Agudo}, {Jorstad}, {Marscher}, {Larionov},
  {G{\'o}mez}, {L{\"a}hteenm{\"a}ki}, {Gurwell}, {Smith}, {Wiesemeyer}, {Thum},
  {Heidt}, {Blinov}, {D'Arcangelo}, {Hagen-Thorn}, {Morozova}, {Nieppola},
  {Roca-Sogorb}, {Schmidt}, {Taylor}, {Tornikoski}, \& {Troitsky}}]{Agudo11}
{Agudo}, I., {Jorstad}, S.~G., {Marscher}, A.~P., {et~al.} 2011, \apjl, 726,
  L13

\bibitem[{{Antonucci}(1984)}]{Antonucci84}
{Antonucci}, R.~R.~J. 1984, \apj, 278, 499

\bibitem[{{Baldwin} \& {Netzer}(1978)}]{Baldwin78}
{Baldwin}, J.~A., \& {Netzer}, H. 1978, \apj, 226, 1

\bibitem[{{Ben{\'{\i}}tez} {et~al.}(2010){Ben{\'{\i}}tez}, {Chavushyan},
  {Raiteri}, {Villata}, {Dultzin}, {Mart{\'{\i}}nez}, {P{\'e}rez-Camargo}, \&
  {Torrealba}}]{Benitez10}
{Ben{\'{\i}}tez}, E., {Chavushyan}, V.~H., {Raiteri}, C.~M., {et~al.} 2010, in
  Astronomical Society of the Pacific Conference Series, Vol. 427, Accretion
  and Ejection in AGN: a Global View, ed. L.~{Maraschi}, G.~{Ghisellini},
  R.~{Della Ceca}, \& F.~{Tavecchio}, 291

\bibitem[{{Bessell} {et~al.}(1998){Bessell}, {Castelli}, \& {Plez}}]{Bessell98}
{Bessell}, M.~S., {Castelli}, F., \& {Plez}, B. 1998, \aap, 333, 231

\bibitem[{{Blandford} \& {Payne}(1982)}]{BlandfordPayne82}
{Blandford}, R.~D., \& {Payne}, D.~G. 1982, \mnras, 199, 883

\bibitem[{{Blandford} \& {Znajek}(1977)}]{BlandfordZnajek77}
{Blandford}, R.~D., \& {Znajek}, R.~L. 1977, \mnras, 179, 433

\bibitem[{{Bonning} {et~al.}(2012){Bonning}, {Urry}, {Bailyn}, {Buxton},
  {Chatterjee}, {Coppi}, {Fossati}, {Isler}, \& {Maraschi}}]{Bonning12}
{Bonning}, E., {Urry}, C.~M., {Bailyn}, C., {et~al.} 2012, \apj, 756, 13

\bibitem[{{Bonning} {et~al.}(2009){Bonning}, {Bailyn}, {Urry}, {Buxton},
  {Fossati}, {Maraschi}, {Coppi}, {Scalzo}, {Isler}, \& {Kaptur}}]{Bonning09}
{Bonning}, E.~W., {Bailyn}, C., {Urry}, C.~M., {et~al.} 2009, \apjl, 697, L81

\bibitem[{{Bonnoli} {et~al.}(2011){Bonnoli}, {Ghisellini}, {Foschini},
  {Tavecchio}, \& {Ghirlanda}}]{Bonnoli11}
{Bonnoli}, G., {Ghisellini}, G., {Foschini}, L., {Tavecchio}, F., \&
  {Ghirlanda}, G. 2011, \mnras, 410, 368

\bibitem[{{B{\"o}ttcher}(2007)}]{Bottcher07}
{B{\"o}ttcher}, M. 2007, \apss, 309, 95

\bibitem[{{Bregman} {et~al.}(1986){Bregman}, {Glassgold}, {Huggins}, \&
  {Kinney}}]{Bregman86}
{Bregman}, J.~N., {Glassgold}, A.~E., {Huggins}, P.~J., \& {Kinney}, A.~L.
  1986, \apj, 301, 698

\bibitem[{{Celotti} {et~al.}(1997){Celotti}, {Padovani}, \&
  {Ghisellini}}]{Celotti97}
{Celotti}, A., {Padovani}, P., \& {Ghisellini}, G. 1997, \mnras, 286, 415

\bibitem[{{Chen} {et~al.}(2009){Chen}, {Gu}, {Fan}, \& {Cao}}]{Chen09}
{Chen}, Z.-Y., {Gu}, M.-F., {Fan}, Z.-H., \& {Cao}, X.-W. 2009, Research in
  Astronomy and Astrophysics, 9, 1192

\bibitem[{{Clavel} {et~al.}(1991){Clavel}, {Reichert}, {Alloin}, {Crenshaw},
  {Kriss}, {Krolik}, {Malkan}, {Netzer}, {Peterson}, {Wamsteker}, {Altamore},
  {Baribaud}, {Barr}, {Beck}, {Binette}, {Bromage}, {Brosch}, {Diaz},
  {Filippenko}, {Fricke}, {Gaskell}, {Giommi}, {Glass}, {Gondhalekar},
  {Hackney}, {Halpern}, {Hutter}, {Joersaeter}, {Kinney}, {Kollatschny},
  {Koratkar}, {Korista}, {Laor}, {Lasota}, {Leibowitz}, {Maoz}, {Martin},
  {Mazeh}, {Meurs}, {Nair}, {O'Brien}, {Pelat}, {Perez}, {Perola}, {Ptak},
  {Rodriguez-Pascual}, {Rosenblatt}, {Sadun}, {Santos-Lleo}, {Shaw}, {Smith},
  {Stirpe}, {Stoner}, {Sun}, {Ulrich}, {van Groningen}, \& {Zheng}}]{Clavel91}
{Clavel}, J., {Reichert}, G.~A., {Alloin}, D., {et~al.} 1991, \apj, 366, 64

\bibitem[{{Coppi}(1992)}]{Coppi92}
{Coppi}, P.~S. 1992, \mnras, 258, 657

\bibitem[{{Corbett} {et~al.}(2000){Corbett}, {Robinson}, {Axon}, \&
  {Young}}]{Corbett00}
{Corbett}, E.~A., {Robinson}, A., {Axon}, D.~J., \& {Young}, S. 2000, \mnras,
  319, 685

\bibitem[{{D'Ammando} {et~al.}(2009){D'Ammando}, {Pucella}, {Raiteri},
  {Villata}, {Vittorini}, {Vercellone}, {Donnarumma}, {Longo}, {Tavani},
  {Argan}, {Barbiellini}, {Boffelli}, {Bulgarelli}, {Caraveo}, {Cattaneo},
  {Chen}, {Cocco}, {Costa}, {Del Monte}, {de Paris}, {Di Cocco}, {Evangelista},
  {Feroci}, {Ferrari}, {Fiorini}, {Froysland}, {Fuschino}, {Galli}, {Gianotti},
  {Giuliani}, {Labanti}, {Lapshov}, {Lazzarotto}, {Lipari}, {Marisaldi},
  {Mereghetti}, {Morselli}, {Pacciani}, {Pellizzoni}, {Perotti}, {Piano},
  {Picozza}, {Pilia}, {Prest}, {Rapisarda}, {Rappoldi}, {Sabatini}, {Soffitta},
  {Trifoglio}, {Trois}, {Vallazza}, {Zambra}, {Zanello}, {Agudo}, {Aller},
  {Aller}, {Arkharov}, {Bach}, {Benitez}, {Berdyugin}, {Blinov}, {Buemi},
  {Chen}, {di Paola}, {di Rico}, {Dultzin}, {Fuhrmann}, {G{\'o}mez}, {Gurwell},
  {Jorstad}, {Heidt}, {Hiriart}, {Hsiao}, {Kimeridze}, {Konstantinova},
  {Kopatskaya}, {Koptelova}, {Kurtanidze}, {Larionov}, {Leto}, {Lindfors},
  {Lopez}, {Marscher}, {McHardy}, {Melnichuk}, {Mommert}, {Mujica}, {Nilsson},
  {Pasanen}, {Roca-Sogorb}, {Sorcia}, {Takalo}, {Taylor}, {Trigilio},
  {Troitsky}, {Umana}, {Antonelli}, {Colafrancesco}, {Cutini}, {Gasparrini},
  {Pittori}, {Preger}, {Santolamazza}, {Verrecchia}, {Giommi}, \&
  {Salotti}}]{DAmmando09}
{D'Ammando}, F., {Pucella}, G., {Raiteri}, C.~M., {et~al.} 2009, \aap, 508, 181

\bibitem[{{Falomo} {et~al.}(1994){Falomo}, {Scarpa}, \&
  {Bersanelli}}]{Falomo94}
{Falomo}, R., {Scarpa}, R., \& {Bersanelli}, M. 1994, \apjs, 93, 125

\bibitem[{{Fossati} {et~al.}(1998){Fossati}, {Maraschi}, {Celotti}, {Comastri},
  \& {Ghisellini}}]{Fossati98}
{Fossati}, G., {Maraschi}, L., {Celotti}, A., {Comastri}, A., \& {Ghisellini},
  G. 1998, \mnras, 299, 433

\bibitem[{{Francis} {et~al.}(1991){Francis}, {Hewett}, {Foltz}, {Chaffee},
  {Weymann}, \& {Morris}}]{Francis91}
{Francis}, P.~J., {Hewett}, P.~C., {Foltz}, C.~B., {et~al.} 1991, \apj, 373,
  465

\bibitem[{{Gaskell} {et~al.}(1981){Gaskell}, {Shields}, \&
  {Wampler}}]{Gaskell81}
{Gaskell}, C.~M., {Shields}, G.~A., \& {Wampler}, E.~J. 1981, \apj, 249, 443

\bibitem[{{Ghisellini} {et~al.}(2007){Ghisellini}, {Foschini}, {Tavecchio}, \&
  {Pian}}]{Ghisellini07}
{Ghisellini}, G., {Foschini}, L., {Tavecchio}, F., \& {Pian}, E. 2007, \mnras,
  382, L82

\bibitem[{{Ghisellini} \& {Madau}(1996)}]{Ghisellini96}
{Ghisellini}, G., \& {Madau}, P. 1996, \mnras, 280, 67

\bibitem[{{Ghisellini} {et~al.}(1993){Ghisellini}, {Padovani}, {Celotti}, \&
  {Maraschi}}]{Ghisellini93}
{Ghisellini}, G., {Padovani}, P., {Celotti}, A., \& {Maraschi}, L. 1993, \apj,
  407, 65

\bibitem[{{Ghisellini} {et~al.}(2011){Ghisellini}, {Tavecchio}, {Foschini}, \&
  {Ghirlanda}}]{Ghisellini11}
{Ghisellini}, G., {Tavecchio}, F., {Foschini}, L., \& {Ghirlanda}, G. 2011,
  \mnras, 414, 2674

\bibitem[{{Ghisellini} {et~al.}(2010){Ghisellini}, {Tavecchio}, {Foschini},
  {Ghirlanda}, {Maraschi}, \& {Celotti}}]{Ghisellini10}
{Ghisellini}, G., {Tavecchio}, F., {Foschini}, L., {et~al.} 2010, \mnras, 402,
  497

\bibitem[{{Goad} {et~al.}(1999){Goad}, {Koratkar}, {Kim-Quijano}, {Korista},
  {O'Brien}, \& {Axon}}]{Goad99}
{Goad}, M.~R., {Koratkar}, A.~P., {Kim-Quijano}, J., {et~al.} 1999, \apj, 524,
  707

\bibitem[{{Jorstad} {et~al.}(2005){Jorstad}, {Marscher}, {Lister}, {Stirling},
  {Cawthorne}, {Gear}, {G{\'o}mez}, {Stevens}, {Smith}, {Forster}, \&
  {Robson}}]{Jorstad05}
{Jorstad}, S.~G., {Marscher}, A.~P., {Lister}, M.~L., {et~al.} 2005, \aj, 130,
  1418

\bibitem[{{Jorstad} {et~al.}(2010){Jorstad}, {Marscher}, {Larionov}, {Agudo},
  {Smith}, {Gurwell}, {L{\"a}hteenm{\"a}ki}, {Tornikoski}, {Markowitz},
  {Arkharov}, {Blinov}, {Chatterjee}, {D'Arcangelo}, {Falcone}, {G{\'o}mez},
  {Hagen-Thorn}, {Jordan}, {Kimeridze}, {Konstantinova}, {Kopatskaya},
  {Kurtanidze}, {Larionova}, {Larionova}, {McHardy}, {Melnichuk},
  {Roca-Sogorb}, {Schmidt}, {Skiff}, {Taylor}, {Thum}, {Troitsky}, \&
  {Wiesemeyer}}]{Jorstad10}
{Jorstad}, S.~G., {Marscher}, A.~P., {Larionov}, V.~M., {et~al.} 2010, \apj,
  715, 362

\bibitem[{{Jorstad} {et~al.}(2012){Jorstad}, {Marscher}, {Joshi}, {MacDonald},
  {Scott}, {Williamson}, {Smith}, {Larionov}, {Agudo}, \&
  {Gurwell}}]{Jorstad12}
{Jorstad}, S.~G., {Marscher}, A.~P., {Joshi}, M., {et~al.} 2012, in, 1st. Fermi
  \& Jansky Symposium Proceedings, eConf Proceedings C1111101.
  (arXiv:1205.0520)

\bibitem[{{Kaspi} {et~al.}(2005){Kaspi}, {Maoz}, {Netzer}, {Peterson},
  {Vestergaard}, \& {Jannuzi}}]{Kaspi05}
{Kaspi}, S., {Maoz}, D., {Netzer}, H., {et~al.} 2005, \apj, 629, 61

\bibitem[{{Kataoka} {et~al.}(2008){Kataoka}, {Madejski}, {Sikora}, {Roming},
  {Chester}, {Grupe}, {Tsubuku}, {Sato}, {Kawai}, {Tosti}, {Impiombato},
  {Kovalev}, {Kovalev}, {Edwards}, {Wagner}, {Moderski}, {Stawarz},
  {Takahashi}, \& {Watanabe}}]{Kataoka08}
{Kataoka}, J., {Madejski}, G., {Sikora}, M., {et~al.} 2008, \apj, 672, 787

\bibitem[{{Kellermann} {et~al.}(1989){Kellermann}, {Sramek}, {Schmidt},
  {Shaffer}, \& {Green}}]{Kellermann89}
{Kellermann}, K.~I., {Sramek}, R., {Schmidt}, M., {Shaffer}, D.~B., \& {Green},
  R. 1989, \aj, 98, 1195

\bibitem[{{Koratkar} {et~al.}(1998){Koratkar}, {Pian}, {Urry}, \&
  {Pesce}}]{Koratkar98}
{Koratkar}, A., {Pian}, E., {Urry}, C.~M., \& {Pesce}, J.~E. 1998, \apj, 492,
  173

\bibitem[{{Korista} {et~al.}(1997){Korista}, {Baldwin}, {Ferland}, \&
  {Verner}}]{Korista97}
{Korista}, K., {Baldwin}, J., {Ferland}, G., \& {Verner}, D. 1997, \apjs, 108,
  401

\bibitem[{{Le{\'o}n-Tavares} {et~al.}(2013){Le{\'o}n-Tavares}, {Chavushyan},
  {Pati{\~n}o-{\'A}lvarez}, {Valtaoja}, {Arshakian}, {Popovi{\'c}},
  {Tornikoski}, {Lobanov}, {Carrami{\~n}ana}, {Carrasco}, \&
  {L{\"a}hteenm{\"a}ki}}]{LT13}
{Le{\'o}n-Tavares}, J., {Chavushyan}, V., {Pati{\~n}o-{\'A}lvarez}, V.,
  {et~al.} 2013, \apjl, 763, L36

\bibitem[{{Maraschi} {et~al.}(1992){Maraschi}, {Ghisellini}, \&
  {Celotti}}]{Maraschi92}
{Maraschi}, L., {Ghisellini}, G., \& {Celotti}, A. 1992, \apjl, 397, L5

\bibitem[{{Markwardt}(2009)}]{Markwardt09}
{Markwardt}, C.~B. 2009, in Astronomical Society of the Pacific Conference
  Series, Vol. 411, Astronomical Data Analysis Software and Systems XVIII, ed.
  D.~A. {Bohlender}, D.~{Durand}, \& P.~{Dowler}, 251

\bibitem[{{Marscher} {et~al.}(2011){Marscher}, {Jorstad}, {Larionov}, {Aller},
  \& {L{\"a}hteenm{\"a}ki}}]{Marscher11}
{Marscher}, A., {Jorstad}, S.~G., {Larionov}, V.~M., {Aller}, M.~F., \&
  {L{\"a}hteenm{\"a}ki}, A. 2011, Journal of Astrophysics and Astronomy, 32,
  233

\bibitem[{{Marscher} {et~al.}(2008){Marscher}, {Jorstad}, {D'Arcangelo},
  {Smith}, {Williams}, {Larionov}, {Oh}, {Olmstead}, {Aller}, {Aller},
  {McHardy}, {L{\"a}hteenm{\"a}ki}, {Tornikoski}, {Valtaoja}, {Hagen-Thorn},
  {Kopatskaya}, {Gear}, {Tosti}, {Kurtanidze}, {Nikolashvili}, {Sigua},
  {Miller}, \& {Ryle}}]{Marscher08}
{Marscher}, A.~P., {Jorstad}, S.~G., {D'Arcangelo}, F.~D., {et~al.} 2008, \nat,
  452, 966

\bibitem[{{Mattox} {et~al.}(1996){Mattox}, {Bertsch}, {Chiang}, {Dingus},
  {Digel}, {Esposito}, {Fierro}, {Hartman}, {Hunter}, {Kanbach}, {Kniffen},
  {Lin}, {Macomb}, {Mayer-Hasselwander}, {Michelson}, {von Montigny},
  {Mukherjee}, {Nolan}, {Ramanamurthy}, {Schneid}, {Sreekumar}, {Thompson}, \&
  {Willis}}]{Mattox96}
{Mattox}, J.~R., {Bertsch}, D.~L., {Chiang}, J., {et~al.} 1996, \apj, 461, 396

\bibitem[{{Netzer} \& {Peterson}(1997)}]{Netzer97}
{Netzer}, H., \& {Peterson}, B.~M. 1997, in Astrophysics and Space Science
  Library, Vol. 218, Astronomical Time Series, ed. D.~{Maoz}, A.~{Sternberg},
  \& E.~M. {Leibowitz}, 85

\bibitem[{{Pacciani} {et~al.}(2010){Pacciani}, {Vittorini}, {Tavani},
  {Fiocchi}, {Vercellone}, {D'Ammando}, {Sakamoto}, {Pian}, {Raiteri},
  {Villata}, {Sasada}, {Itoh}, {Yamanaka}, {Uemura}, {Striani}, {Fugazza},
  {Tiengo}, {Krimm}, {Stroh}, {Falcone}, {Curran}, {Sadun}, {Lahteenmaki},
  {Tornikoski}, {Aller}, {Aller}, {Lin}, {Larionov}, {Leto}, {Takalo},
  {Berdyugin}, {Gurwell}, {Bulgarelli}, {Chen}, {Donnarumma}, {Giuliani},
  {Longo}, {Pucella}, {Argan}, {Barbiellini}, {Caraveo}, {Cattaneo}, {Costa},
  {De Paris}, {Del Monte}, {Di Cocco}, {Evangelista}, {Ferrari}, {Feroci},
  {Fiorini}, {Fuschino}, {Galli}, {Gianotti}, {Labanti}, {Lapshov},
  {Lazzarotto}, {Lipari}, {Marisaldi}, {Mereghetti}, {Morelli}, {Moretti},
  {Morselli}, {Pellizzoni}, {Perotti}, {Piano}, {Picozza}, {Pilia}, {Prest},
  {Rapisarda}, {Rappoldi}, {Rubini}, {Sabatini}, {Soffitta}, {Trifoglio},
  {Trois}, {Vallazza}, {Zanello}, {Colafrancesco}, {Pittori}, {Verrecchia},
  {Santolamazza}, {Lucarelli}, {Giommi}, \& {Salotti}}]{Pacciani10}
{Pacciani}, L., {Vittorini}, V., {Tavani}, M., {et~al.} 2010, \apjl, 716, L170

\bibitem[{{Perez} {et~al.}(1989){Perez}, {Penston}, \& {Moles}}]{Perez89}
{Perez}, E., {Penston}, M.~V., \& {Moles}, M. 1989, \mnras, 239, 75

\bibitem[{{Peterson}(1993)}]{Peterson93}
{Peterson}, B.~M. 1993, \pasp, 105, 247

\bibitem[{{Peterson}(2006)}]{Peterson06}
{Peterson}, B.~M. 2006, in Astronomical Society of the Pacific Conference
  Series, Vol. 360, Astronomical Society of the Pacific Conference Series, ed.
  C.~M. {Gaskell}, I.~M. {McHardy}, B.~M. {Peterson}, \& S.~G. {Sergeev}, 191

\bibitem[{{Peterson} \& {Ferland}(1986)}]{Peterson86}
{Peterson}, B.~M., \& {Ferland}, G.~J. 1986, \nat, 324, 345

\bibitem[{{Pian} {et~al.}(1998){Pian}, {Koratkar}, {Maraschi}, {Urry},
  {Madejsi}, {McHardy}, {Pesce}, {Treves}, {Grandi}, \& {Leach}}]{Pian98}
{Pian}, E., {Koratkar}, A., {Maraschi}, L., {et~al.} 1998, in ESA Special
  Publication, Vol. 413, Ultraviolet Astrophysics Beyond the IUE Final Archive,
  ed. W.~{Wamsteker}, R.~{Gonzalez Riestra}, \& B.~{Harris}, 615

\bibitem[{{Poutanen} \& {Stern}(2010)}]{Poutanen10}
{Poutanen}, J., \& {Stern}, B. 2010, \apjl, 717, L118

\bibitem[{{Raiteri} {et~al.}(2007{\natexlab{a}}){Raiteri}, {Villata},
  {Capetti}, {Heidt}, {Arnaboldi}, \& {Magazz{\`u}}}]{Raiteri07_0235}
{Raiteri}, C.~M., {Villata}, M., {Capetti}, A., {et~al.} 2007{\natexlab{a}},
  \aap, 464, 871

\bibitem[{{Raiteri} {et~al.}(2007{\natexlab{b}}){Raiteri}, {Villata},
  {Larionov}, {Pursimo}, {Ibrahimov}, {Nilsson}, {Aller}, {Kurtanidze},
  {Foschini}, {Ohlert}, {Papadakis}, {Sumitomo}, {Volvach}, {Aller},
  {Arkharov}, {Bach}, {Berdyugin}, {B{\"o}ttcher}, {Buemi}, {Calcidese},
  {Charlot}, {Delgado S{\'a}nchez}, {di Paola}, {Djupvik}, {Dolci}, {Efimova},
  {Fan}, {Forn{\'e}}, {Gomez}, {Gupta}, {Hagen-Thorn}, {Hooks}, {Hovatta},
  {Ishii}, {Kamada}, {Konstantinova}, {Kopatskaya}, {Kovalev}, {Kovalev},
  {L{\"a}hteenm{\"a}ki}, {Lanteri}, {Le Campion}, {Lee}, {Leto}, {Lin},
  {Lindfors}, {Mingaliev}, {Mizoguchi}, {Nicastro}, {Nikolashvili},
  {Nishiyama}, {{\"O}stman}, {Ovcharov}, {P{\"a}{\"a}kk{\"o}nen}, {Pasanen},
  {Pian}, {Rector}, {Ros}, {Sadakane}, {Selj}, {Semkov}, {Sharapov}, {Somero},
  {Stanev}, {Strigachev}, {Takalo}, {Tanaka}, {Tavani}, {Torniainen},
  {Tornikoski}, {Trigilio}, {Umana}, {Vercellone}, {Valcheva}, {Volvach}, \&
  {Yamanaka}}]{Raiteri07}
{Raiteri}, C.~M., {Villata}, M., {Larionov}, V.~M., {et~al.}
  2007{\natexlab{b}}, \aap, 473, 819

\bibitem[{{Raiteri} {et~al.}(2011){Raiteri}, {Villata}, {Aller}, {Gurwell},
  {Kurtanidze}, {L{\"a}hteenm{\"a}ki}, {Larionov}, {Romano}, {Vercellone},
  {Agudo}, {Aller}, {Arkharov}, {Bach}, {Ben{\'{\i}}tez}, {Berdyugin},
  {Blinov}, {Borisova}, {B{\"o}ttcher}, {Bravo Calle}, {Buemi}, {Calcidese},
  {Carosati}, {Casas}, {Chen}, {Efimova}, {G{\'o}mez}, {Gusbar}, {Hawkins},
  {Heidt}, {Hiriart}, {Hsiao}, {Jordan}, {Jorstad}, {Joshi}, {Kimeridze},
  {Koptelova}, {Konstantinova}, {Kopatskaya}, {Kurtanidze}, {Larionova},
  {Larionova}, {Leto}, {Li}, {Ligustri}, {Lindfors}, {Lister}, {Marscher},
  {Molina}, {Morozova}, {Nieppola}, {Nikolashvili}, {Nilsson}, {Palma},
  {Pasanen}, {Reinthal}, {Roberts}, {Ros}, {Roustazadeh}, {Sadun}, {Sakamoto},
  {Schwartz}, {Sigua}, {Sillanp{\"a}{\"a}}, {Takalo}, {Tammi}, {Taylor},
  {Tornikoski}, {Trigilio}, {Troitsky}, {Umana}, {Volvach}, \&
  {Yuldasheva}}]{Raiteri11}
{Raiteri}, C.~M., {Villata}, M., {Aller}, M.~F., {et~al.} 2011, \aap, 534, A87

\bibitem[{{Raiteri} {et~al.}(2012){Raiteri}, {Villata}, {Smith}, {Larionov},
  {Acosta-Pulido}, {Aller}, {D'Ammando}, {Gurwell}, {Jorstad}, {Joshi},
  {Kurtanidze}, {L{\"a}hteenm{\"a}ki}, {Mirzaqulov}, {Agudo}, {Aller},
  {Ar{\'e}valo}, {Arkharov}, {Bach}, {Ben{\'{\i}}tez}, {Berdyugin}, {Blinov},
  {Blumenthal}, {Buemi}, {Bueno}, {Carleton}, {Carnerero}, {Carosati},
  {Casadio}, {Chen}, {Di Paola}, {Dolci}, {Efimova}, {Ehgamberdiev},
  {G{\'o}mez}, {Gonz{\'a}lez}, {Hagen-Thorn}, {Heidt}, {Hiriart}, {Holikov},
  {Konstantinova}, {Kopatskaya}, {Koptelova}, {Kurtanidze}, {Larionova},
  {Larionova}, {Le{\'o}n-Tavares}, {Leto}, {Lin}, {Lindfors}, {Marscher},
  {McHardy}, {Molina}, {Morozova}, {Mujica}, {Nikolashvili}, {Nilsson},
  {Ovcharov}, {Panwar}, {Pasanen}, {Puerto-Gimenez}, {Reinthal}, {Richter},
  {Ros}, {Sakamoto}, {Schwartz}, {Sillanp{\"a}{\"a}}, {Smith}, {Takalo},
  {Tammi}, {Taylor}, {Thum}, {Tornikoski}, {Trigilio}, {Troitsky}, {Umana},
  {Valcheva}, \& {Wehrle}}]{Raiteri12}
{Raiteri}, C.~M., {Villata}, M., {Smith}, P.~S., {et~al.} 2012, \aap, 545, A48

\bibitem[{{Sbarrato} {et~al.}(2012){Sbarrato}, {Ghisellini}, {Maraschi}, \&
  {Colpi}}]{Sbarrato12}
{Sbarrato}, T., {Ghisellini}, G., {Maraschi}, L., \& {Colpi}, M. 2012, \mnras,
  421, 1764

\bibitem[{{Shaw} {et~al.}(2012){Shaw}, {Romani}, {Cotter}, {Healey},
  {Michelson}, {Readhead}, {Richards}, {Max-Moerbeck}, {King}, \&
  {Potter}}]{Shaw12}
{Shaw}, M.~S., {Romani}, R.~W., {Cotter}, G., {et~al.} 2012, \apj, 748, 49

\bibitem[{{Smith} {et~al.}(1986){Smith}, {Balonek}, {Heckert}, \&
  {Elston}}]{Smith86}
{Smith}, P.~S., {Balonek}, T.~J., {Heckert}, P.~A., \& {Elston}, R. 1986, \apj,
  305, 484

\bibitem[{{Smith} {et~al.}(1988){Smith}, {Elston}, {Berriman}, {Allen}, \&
  {Balonek}}]{Smith88}
{Smith}, P.~S., {Elston}, R., {Berriman}, G., {Allen}, R.~G., \& {Balonek},
  T.~J. 1988, \apjl, 326, L39

\bibitem[{{Smith} {et~al.}(2009){Smith}, {Montiel}, {Rightley}, {Turner},
  {Schmidt}, \& {Jannuzi}}]{Smith09}
{Smith}, P.~S., {Montiel}, E., {Rightley}, S., {et~al.} 2009, in 2nd. Fermi
  Symposium Proceedings, eConf Proceedings C091122. (arXiv:0912.3621)

\bibitem[{{Smith} {et~al.}(2011){Smith}, {Schmidt}, \& {Jannuzi}}]{Smith11}
{Smith}, P.~S., {Schmidt}, G.~D., \& {Jannuzi}, B.~T. 2011, in, 4th. Fermi
  Symposium Proceedings, eConf Proceedings C110509. (arXiv:1110:6040)

\bibitem[{{Smith} {et~al.}(1994){Smith}, {Schmidt}, {Jannuzi}, \&
  {Elston}}]{Smith94}
{Smith}, P.~S., {Schmidt}, G.~D., {Jannuzi}, B.~T., \& {Elston}, R. 1994, \apj,
  426, 535

\bibitem[{{Stanishev}(2007)}]{Stanishev07}
{Stanishev}, V. 2007, Astronomische Nachrichten, 328, 948

\bibitem[{{Stern} \& {Poutanen}(2011)}]{Stern11}
{Stern}, B.~E., \& {Poutanen}, J. 2011, \mnras, 417, L11

\bibitem[{{Szokoly} {et~al.}(2004){Szokoly}, {Bergeron}, {Hasinger}, {Lehmann},
  {Kewley}, {Mainieri}, {Nonino}, {Rosati}, {Giacconi}, {Gilli}, {Gilmozzi},
  {Norman}, {Romaniello}, {Schreier}, {Tozzi}, {Wang}, {Zheng}, \&
  {Zirm}}]{Szokoly04}
{Szokoly}, G.~P., {Bergeron}, J., {Hasinger}, G., {et~al.} 2004, \apjs, 155,
  271

\bibitem[{{Tadhunter} {et~al.}(1992){Tadhunter}, {Scarrott}, {Draper}, \&
  {Rolph}}]{Tadhunter92}
{Tadhunter}, C.~N., {Scarrott}, S.~M., {Draper}, P., \& {Rolph}, C. 1992,
  \mnras, 256, 53P

\bibitem[{{Tavecchio} {et~al.}(2010){Tavecchio}, {Ghisellini}, {Bonnoli}, \&
  {Ghirlanda}}]{Tavecchio10}
{Tavecchio}, F., {Ghisellini}, G., {Bonnoli}, G., \& {Ghirlanda}, G. 2010,
  \mnras, 405, L94

\bibitem[{{Ulrich} {et~al.}(1993){Ulrich}, {Courvoisier}, \&
  {Wamsteker}}]{Ulrich93}
{Ulrich}, M.-H., {Courvoisier}, T.~J.-L., \& {Wamsteker}, W. 1993, \apj, 411,
  125

\bibitem[{{Ulrich} {et~al.}(1997){Ulrich}, {Maraschi}, \& {Urry}}]{Ulrich97}
{Ulrich}, M.-H., {Maraschi}, L., \& {Urry}, C.~M. 1997, \araa, 35, 445

\bibitem[{{Urry} \& {Padovani}(1995)}]{Urry95}
{Urry}, C.~M., \& {Padovani}, P. 1995, \pasp, 107, 803

\bibitem[{{Vercellone} {et~al.}(2011){Vercellone}, {Striani}, {Vittorini},
  {Donnarumma}, {Pacciani}, {Pucella}, {Tavani}, {Raiteri}, {Villata},
  {Romano}, {Fiocchi}, {Bazzano}, {Bianchin}, {Ferrigno}, {Maraschi}, {Pian},
  {T{\"u}rler}, {Ubertini}, {Bulgarelli}, {Chen}, {Giuliani}, {Longo},
  {Barbiellini}, {Cardillo}, {Cattaneo}, {Del Monte}, {Evangelista}, {Feroci},
  {Ferrari}, {Fuschino}, {Gianotti}, {Giusti}, {Lazzarotto}, {Pellizzoni},
  {Piano}, {Pilia}, {Rapisarda}, {Rappoldi}, {Sabatini}, {Soffitta},
  {Trifoglio}, {Trois}, {Giommi}, {Lucarelli}, {Pittori}, {Santolamazza},
  {Verrecchia}, {Agudo}, {Aller}, {Aller}, {Arkharov}, {Bach}, {Berdyugin},
  {Borman}, {Chigladze}, {Efimov}, {Efimova}, {G{\'o}mez}, {Gurwell},
  {McHardy}, {Joshi}, {Kimeridze}, {Krajci}, {Kurtanidze}, {Kurtanidze},
  {Larionov}, {Lindfors}, {Molina}, {Morozova}, {Nazarov}, {Nikolashvili},
  {Nilsson}, {Pasanen}, {Reinthal}, {Ros}, {Sadun}, {Sakamoto}, {Sallum},
  {Sergeev}, {Schwartz}, {Sigua}, {Sillanp{\"a}{\"a}}, {Sokolovsky},
  {Strelnitski}, {Takalo}, {Taylor}, \& {Walker}}]{Vercellone11}
{Vercellone}, S., {Striani}, E., {Vittorini}, V., {et~al.} 2011, \apjl, 736,
  L38

\bibitem[{{Villata} {et~al.}(2007){Villata}, {Raiteri}, {Aller}, {Bach},
  {Ibrahimov}, {Kovalev}, {Kurtanidze}, {Larionov}, {Lee}, {Leto},
  {L{\"a}hteenm{\"a}ki}, {Nilsson}, {Pursimo}, {Ros}, {Sumitomo}, {Volvach},
  {Aller}, {Arai}, {Buemi}, {Coloma}, {Doroshenko}, {Efimov}, {Fuhrmann},
  {Hagen-Thorn}, {Kamada}, {Katsuura}, {Konstantinova}, {Kopatskaya}, {Kotaka},
  {Kovalev}, {Kurosaki}, {Lanteri}, {Larionova}, {Mingaliev}, {Mizoguchi},
  {Nakamura}, {Nikolashvili}, {Nishiyama}, {Sadakane}, {Sergeev}, {Sigua},
  {Sillanp{\"a}{\"a}}, {Smart}, {Takalo}, {Tanaka}, {Tornikoski}, {Trigilio},
  \& {Umana}}]{Villata07}
{Villata}, M., {Raiteri}, C.~M., {Aller}, M.~F., {et~al.} 2007, \aap, 464, L5,
  knot explanation for blazar variability

\bibitem[{{Wehrle} {et~al.}(2012){Wehrle}, {Marscher}, {Jorstad}, {Gurwell},
  {Joshi}, {MacDonald}, {Williamson}, {Agudo}, \& {Grupe}}]{Wehrle12}
{Wehrle}, A.~E., {Marscher}, A.~P., {Jorstad}, S.~G., {et~al.} 2012, \apj, 758,
  72

\bibitem[{{Wills} {et~al.}(1985){Wills}, {Netzer}, \& {Wills}}]{Wills85}
{Wills}, B.~J., {Netzer}, H., \& {Wills}, D. 1985, \apj, 288, 94

\bibitem[{{Zheng} \& {Burbidge}(1986)}]{Zheng86}
{Zheng}, W., \& {Burbidge}, E.~M. 1986, \apjl, 306, L67

\end{thebibliography}

\end{document}